\documentclass[fleqn,usenatbib]{mnras}

\usepackage{newtxtext,newtxmath}

\usepackage[T1]{fontenc}

\DeclareRobustCommand{\VAN}[3]{#2}
\let\VANthebibliography\thebibliography
\def\thebibliography{\DeclareRobustCommand{\VAN}[3]{##3}\VANthebibliography}

\usepackage{graphicx}	
\usepackage{amsmath}	
\usepackage{hyperref}
\usepackage{longtable}
\usepackage{dcolumn}
\usepackage{array}
\usepackage{threeparttable}
\usepackage{tabularx}
\usepackage{lscape}
\usepackage{pdflscape}
\usepackage{multirow}
\usepackage{soul,color}	
\usepackage{xcolor}
\usepackage{adjustbox}
\usepackage{caption}

\newcommand{\kms}{km\,s$^{-1}$}                             
\newcommand{\Gaia}{{\slshape{Gaia}}}            

\title[Photometric variability of UCDs with TESS]{Exploring the photometric variability of ultra-cool dwarfs with TESS}

\author[R. P. Petrucci et al.]{
Romina P. Petrucci,$^{1, 2}$\thanks{E-mail: romina.petrucci@unc.edu.ar}
Yilen G\'omez Maqueo Chew,$^{3}$
Emiliano Jofr\'e,$^{1,2}$
Ant\'igona Segura,$^{4}$
\newauthor
and Leticia V. Ferrero$^{1}$
\\
\\
$^{1}$Universidad Nacional de Córdoba, Observatorio Astronómico de Córdoba, Laprida 854, X5000BGR Córdoba, Argentina\\
$^{2}$Consejo Nacional de Investigaciones Científicas y Técnicas (CONICET), Godoy Cruz 2290, CABA, CPC 1425FQB, Argentina\\
$^{3}$Instituto de Astronomía, Universidad Nacional Autónoma de México, Circuito Exterior, C.U., A. Postal 70-264, 04510 Ciudad de México, México\\
$^{4}$Instituto de Ciencias Nucleares, Universidad Nacional Autónoma de México, Cto. Exterior S/N, C.U., Coyoacán, 04510 Ciudad de Mexico, México
}

\date{Accepted XXX. Received YYY; in original form ZZZ}

\pubyear{2023}

\begin{document}
\label{firstpage}
\pagerange{\pageref{firstpage}--\pageref{lastpage}}
\maketitle

\begin{abstract}
We present a photometric characterization of 208 ultra-cool dwarfs (UCDs) with spectral types between M4 and L4, from 20-second and 2-minute cadence TESS light curves. We determine rotation periods for 87 objects ($\sim42\%$) and identify 778 flare events in 103 UCDs ($\sim49.5\%$). 
For 777 flaring events (corresponding to 102 objects), we derive bolometric energies between $2.1 \times 10^{30}$ and $1.1 \times 10^{34} \mathrm{erg\,}$, with 56 superflare events.
No transiting planets or eclipsing binaries were identified.
We find that the fraction of UCDs with rotation and flaring activity is, at least, 20$\%$ higher in M4--M6 spectral types than in later UCDs (M7--L4). 
For spectral types between M4 and L0, we measure the slope of the flare bolometric energy-duration correlation to be $\gamma = 0.497 \pm 0.058$, which agrees with that found in previous studies for solar-type and M dwarfs.
Moreover, we determine the slope of the flare frequency distribution to be $\alpha = \mathrm{-1.75 \pm 0.04}$ for M4--M5 dwarfs, $\alpha = \mathrm{-1.69 \pm 0.04}$ and $\alpha = \mathrm{-1.72 \pm 0.1}$ for M6--M7 and M8--L0 dwarfs, respectively, which are consistent with
previous works that exclusively analysed UCDs. These results support the idea that independently of the physical mechanisms that produce magnetic activity, the characteristics of the rotational modulation and flares are similar for both fully-convective UCDs and partially-convective solar-type and early-M stars. 
Based on the measured UCD flare distributions, we find that UV radiation emitted from flares does not have the potential to start prebiotic chemistry.

\end{abstract}




\section{Introduction}

Ultra-cool dwarfs (UCDs) are objects with effective temperatures below 3000 K that include fully-convective very-low mass stars and brown
dwarfs \citep[e.g.][]{kirkpatick1995, bolmont2017}. They are particularly interesting because it is easier and more likely to detect Earth-like planets in the habitable zone than around stars of any other spectral type \citep{scalo2007}. However, a key aspect to assess whether (or not) planets orbiting UCDs would be able to sustain life on their surfaces, is to 
characterize the host's magnetic activity. 
In solar-type stars, magnetic activity is described by an $\alpha\omega$ dynamo \citep{parker1955, charbonneau2010} powered by the interaction between stellar rotation and convection. It is believed that the tachocline, i.e. the transition zone between the radiative core that rotates as a solid body and the convective envelope that presents differential rotation, is where the magnetic field organizes and amplifies and, hence, a fundamental element in the dynamo mechanism. Nevertheless, fully-convective stars do not possess this interface, but
magnetic fields of the order of a few kiloGauss have been measured in these objects \citep[see][for a review]{kochukhov2021}. 
Moreover, in a recent study, \cite{climent2023} reported spatially resolved radio observations of a brown dwarf, which were attributed to a dipole-ordered magnetic field with a radiation belt-like morphology.
It means that these stars should also harbor a magnetic dynamo although different from that in solar-type stars. Several models have been proposed \citep[e.g.][]{chabrier2006, browning2008, gastine2013}, however, the underlying mechanism which creates and sustains the magnetic fields in fully-convective stars remains unknown. 

In this scenario, it becomes important to determine if the signatures of this magnetic activity, such as rotation periods and flares, in fully-convective stars follow the same correlations and have similar characteristics than in stars with a radiative core. 
From an evolutionary point of view, during the pre-main sequence phase, both partially- and fully-convective stars are known to exhibit evidence of magnetic activity such as cool starspots \citep[e.g.][]{bouvier2007}, energetic flares \citep[e.g.][]{cody2022, rebull2022}, and high surface magnetic fields \citep[e.g.][]{flores2019, lopezvaldivia2023}, powered by strong magnetic dynamos. In this context, several monitoring campaigns have been launched with the pursuit of exploring the periodic and non-periodic variability of stars of different masses in forming regions and young clusters \citep[e.g.][]{bouvier2007, cody2010, serna2021, getman2022, getman2023}. These previous studies have revealed differences in the rotational properties of stars with distinct masses. Stars with spectral types earlier than M2.5, sometimes present a bimodal distribution with rotation periods predominantly at $\sim 2$ and $\sim 10$\,days, whilst later type objects present a single-peak distribution with rotation periods between $\sim 1$ and $\sim 3$\,days \citep{herbst2002, lamm2005}.
During the main-sequence phase, it is well established that the equatorial rotation speed of FGK and early-M stars declines with the inverse square root of the star's age \citep{skumanich1972} due to angular momentum loss driven by magnetized stellar winds \citep[e.g.][]{angus2020, metcalfe2023}. Nonetheless, this seems not be the case for fully-convective low-mass stars \citep[e.g.][]{tannock2021}.  

Moreover, one relevant parameter to understand the functioning of the magnetic dynamo on 
fully-convective stars is the slope of the flare frequency distribution, $\alpha$, which provides information about how flares yield the magnetic energy responsible for the heating of the corona \citep{parker1955}. In this sense, two recent works \citep{seli2021, murray} that exclusively analyse UCDs
obtained their flare frequency distributions and found slopes of $\alpha$ $\sim -$2, consistent with the range observed for FGK and early-M stars  \citep[e.g.][]{tu2020, gunther2020, jackman2021, yang2023}.
Regarding stellar rotation-activity relationship, previous works \citep{wright2018, newton2017, medina2022} have shown no distinction between stars with and without tachocline. Additionally, some studies that quantified
the correlation between duration and bolometric energy of flares in M stars \citep{silverberg2016, yang2023}, agree with the results obtained for solar-type stars.

In this context, the main purpose of this study is to provide some insight
through the exploration of the photometric variability of a sample of mid-to-late M dwarfs with 2-minute cadence TESS data. The article is organized as follows. In section \ref{observations}, we introduce the sample with their main properties and describe the observational data. The methodology applied to search for rotational modulation, flares and hints of planetary candidates is detailed in section \ref{methods}. In section \ref{results}, we describe our results regarding the search for correlations between rotation and flare's parameters, the galactic kinematics of UCDs, the construction of the flare frequency distributions, and the identification of correlations between amplitude, duration, and energy of flares. Here, we also present our findings about superflares, and briefly 
assess the habitability around UCDs. Finally, we present our conclusions in section \ref{conclusions}.

\section{Sample selection and observations}\label{observations}

Our sample comprises a total of 208 UCDs with spectral types from M4 to L4, extracted from the catalog of M and L dwarfs within 40 pc of \cite{sebastian2021}. For our
sample selection, we first choose targets from `programme 1' of the catalog, which consists of 365 late-type objects that are sufficiently small and close to allow a detailed atmospheric characterization of an hypotetical gravitationally-bounded transiting Earth-like planet with JWST \citep{gardner2006}. Afterward, we performed a cross-match of these data with the TESS Input Catalog \citep[TICv8.2]{stassun2019}, taking as reference the scripts provided on the Mikulski Archive for Space Telescopes (MAST) server\footnote{\url{https://mast.stsci.edu/api/v0/_services.html}}. As a result, from the original `programme 1' list from \citet{sebastian2021}, we kept the 235 targets observed by the Transiting Exoplanet Survey Satellite \citep[TESS]{ricker2015} with 2-minute cadence data available. Concretely, we analysed the Presearch Data Conditioning Simple Aperture Photometry (PDCSAP), processed with the TESS Science Processing Operations Center (SPOC) pipeline \citep{jenkins2016}, with the tools provided by the \textsc{Lightkurve} Python package \citep{lightkurve2018}. Then, given that some targets showed light curves with unphysical
values of flux (i.e. negative values), we had to remove them from the list, reducing our final sample to 208 objects. For each of them, we used the 2-minute cadence data of all the TESS sectors accessible at the time 
of the analysis between sectors 1 and 53. In particular, for those UCDs in our sample with available 20-second light curves, we also used these short-cadence data for a comprehensive study of stellar flares. In summary, the number of targets per spectral type studied in this work is: 6 M4, 61 M5, 64 M6, 29 M7, 15 M8, 14 M9, 10 L0, 5 L1, 2 L2, 1 L3, and 1 L4. In Table \ref{tab_1}, we present their TICv8.2 names and main properties.

\begin{figure}
   \centering
   \includegraphics[width=0.48\textwidth,trim=1.1cm 0.3cm 2cm 1cm, clip]{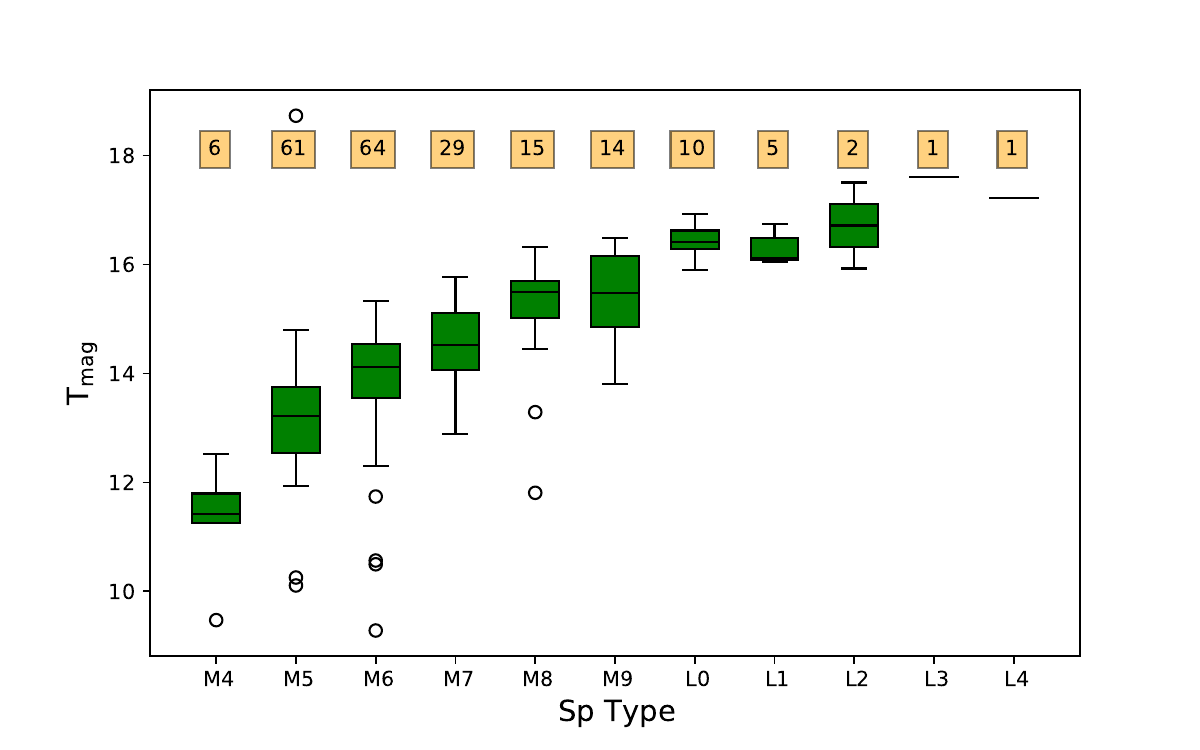}
   \caption{TESS magnitude per spectral type of the 208 UCDs analysed in this work. The horizontal line inside each box indicates the median value for a given spectral type. Outliers are marked as open circles. We note that M4--M5 UCDs are the brightest targets in our sample. Then, T$_{\rm mag}$ increases from M6 to L4 reaching its maximum between L2 and L4 objects. Yellow squares indicate the number of UCDs for each spectral type.}
              \label{tmag}
\end{figure}

In Fig. \ref{tmag}, we show a box plot of the TESS magnitude (T$_{\rm mag}$) extracted from the TESS Input Catalog \citep[TICv8.2]{stassun2019} as a function of spectral type for all the UCDs in the sample. The median T$_{\rm mag}$ value for our full sample is 14.10, with T$_{\rm mag} = 9.28$ and T$_{\rm mag} = 18.73$ for the brightest and faintest objects, respectively. As expected, median T$_{\rm mag}$ values increase from early to late spectral types.   

\clearpage
\begin{landscape}
\begin{table}
\centering
\caption{Main properties of the 208 UCDs analysed in this work.} 
\label{tab_1} 
\begin{threeparttable}
\resizebox{\columnwidth}{!}{
\begin{tabular}{*{25}{c}}
\hline
\noalign{\smallskip}
 \multirow{2}{3em}{TIC ID} & T$_{\rm eff}$ & \multirow{2}{2em}{SpT} &  \multirow{2}{3em}{TESS sectors} & \multirow{2}{3em}{Flares} & \multirow{2}{3em}{Rotation} & \multirow{2}{3em}{$f_{\rm TIC}$} & $\mathrm{L_\mathrm{bol}}$ & P$_{\rm rot}$ & error\_P$_{\rm rot}$ & Amp$^{\textit{a}}$ & error\_Amp & \multirow{2}{3em}{FAP} & U & V & W & {U$_{\rm LSR}$} & {V$_{\rm LSR}$} & {W$_{\rm LSR}$} & eU & eV & eW & \multirow{2}{6em}{RV Reference} & Galactic & \multirow{2}{3em}{Prob.} \\
 & (K) & & & & & & ({erg\,s$^{-1}$}) & (days) & (days) & (mag) & (mag) & & (\kms) & (\kms) & (\kms) & (\kms) & (\kms) & (\kms) & (\kms) & (\kms) & (\kms) & & Population &  \\
\noalign{\smallskip}
\hline
\noalign{\smallskip}
401945077 &	2819 & M6.1	& 5 & No & Yes & 0.00281 & -- & 0.7289 & 0.0204 & 0.0017 & 0.00011 & 4.62 $\times 10^{-32}$ & 25.567 & -23.157 & -6.304 & 35.147 & -12.637 & 0.706 & 0.773 & 0.028 & 0.065 & SIMBAD & THIN-DISK & 0.986\\
298907057 & 2714 & M6.8 & 5,32 & Yes & Yes & 0.00842 & 1.297 $\times 10^{30}$ & 0.5003	& 0.00033 & 0.0096 & 0.00006 & $\ll 1.00 \times 10^{-103}$ & -30.658 & -15.041 & -1.322 & -21.078 & -4.521	& 5.688	& 0.83	& 0.065	& 0.103	& SIMBAD & THIN-DISK & 0.988\\
229115214	& 2864	& M5.4	& 2,29	& Yes & Yes & 0.00043 & 2.930 $\times 10^{30}$	& 2.4843 & 0.00802 & 0.0018 & 0.00009 & 2.77 $\times 10^{-58}$ & 43.504	& -14.55 & 12.411 & 53.084 & -4.03	& 19.421 & 0.22	& 0.012	& 0.01	& SIMBAD & THIN-DISK & 0.973\\
100907328 & 2711 & M6.8 & 30 & No & No & 0.00632 & -- & -- & -- & -- & -- & -- & -40.984 & -19.62 & 11.493 & -31.404 & -9.1 & 18.503 & 0.419 & 0.032 & 0.018 & SIMBAD & THIN-DISK & 0.98\\
63781635 & 2991 & M5 & 18 & Yes & Yes & 0.04523 & 3.851 $\times 10^{30}$ & 0.2761	& 0.00324 & 0.0073 & 0.00001 & $\ll 1.00 \times 10^{-103}$ & -2.121 & 8.195 & -12.319 & 7.459 & 18.715 & -5.309 & 1.867 & 1.795 & 2.064 & SIMBAD & THIN-DISK & 0.988\\
232970271 &	2900 & M5.5 & 14,15,21,22 & Yes & Yes & 0.01921 & 3.258 $\times 10^{30}$ & 0.5105 & 0.00107 & 0.007 & 0.00002 & $\ll 1.00 \times 10^{-103}$ & -29.913	& -16.676 & 0.078 & -20.333 & -6.156 & 7.088 & 0.139 & 0.012 & 0.002 & SIMBAD & THIN-DISK & 0.988\\
187092382 &	2901 & M5.5 & 19 & Yes & Yes & 0.01323 & 3.109 $\times 10^{30}$ & 0.5665 & 0.01277 & 0.0077 & 0.0001 & $\ll 1.00 \times 10^{-103}$	& -4.02	& 11.771 & -11.699	& 5.56 & 22.291 & -4.689 & 2.818 & 2.711 & 1.069 & SIMBAD & THIN-DISK & 0.987\\
24108819  & 2941 & M5.3 & 24,25 & No & No & 0.00585 & -- & -- & -- & -- & -- & -- & -41.689 & -54.238 & 5.331 & -32.109 & -43.718 & 12.341 & 0.597 & 0.023 & 0.028 & SIMBAD & THIN-DISK & 0.981\\
441706467 & 2687 & M7 & 16,22,23,49,50 & No & Yes & 0.00058 & -- & 0.4828 & 0.00107 & 0.0014 & 0.00005 & 6.81 $\times 10^{-28}$ & 7.265 & 42.884 & 8.843 & 16.845 & 53.404 & 15.853 & 0.336 & 0.034 & 0.035 & SIMBAD & THIN-DISK & 0.967\\
286447344 & 2818 & M6.1 & 19 & No & Yes & 0.02116 & -- & 0.7278 & 0.02108 & 0.0046 & 0.00027 & 7.1 $\times 10^{-29}$ & -45.879 & -4.576 & -19.817 & -36.299 & 5.944 & -12.807 & 6.55 & 6.523 & 1.359 & GAIA-DR3 & THIN-DISK & 0.982\\
1042982 & 2807 & M6.2 & 21,48 & Yes & No & 0.0488 & 1.762 $\times 10^{30}$	& -- & -- & -- & -- & -- &	-14.153	& -50.831 & 8.082 & -4.573 &-40.311	& 15.092 & 0.283 & 0.062 & 0.018 & SIMBAD & THIN-DISK & 0.975\\
365064283 & 2745 & M6.6 & 23,46,50 & Yes & No & 0.00279 & 1.798 $\times 10^{30}$ & -- & -- & -- & -- & -- & -13.426 & -13.36 & 9.549 & -3.846 & -2.84 & 16.559 & 0.392 & 0.036 & 0.011 & SIMBAD & THIN-DISK & 0.984\\
87378424 & 2826 & M6 & 19 & No & No & 0.04286 & -- & -- & -- & -- & -- & -- &	-13.879 & -25.186 & -7.33 & -4.299 & -14.666 & -0.32 & 0.478 & 0.02 & 0.006 & SIMBAD & THIN-DISK & 0.988\\
43213934& 2845 & M5.9 & 23 & Yes & Yes & 1.58378 & 2.724 $\times 10^{30}$ & 0.7488 & 0.02276 & 0.0183 & 0.00019 & $\ll 1.00 \times 10^{-103}$	& 9.118 & 6.399 & 2.739 & 18.698 & 16.919 & 9.749 & 0.473 & 0.006 & 0.003 & SIMBAD & THIN-DISK & 0.985\\
17970570 & 2951 & M5.2 & 22,48 & Yes & Yes & 0.00004 & 3.544 $\times 10^{30}$ & 0.5593 & 0.00041 & 0.004 & 0.00004 & 6.48 $\times 10^{-211}$ & 19.989 & 8.545 & 1.055 & 29.569 & 19.065 & 8.065 & 1.546 & 0.084 & 4.671 & SIMBAD & THIN-DISK & 0.983\\
$\hdots$ & $\hdots$ & $\hdots$ & $\hdots$ & $\hdots$ & $\hdots$ & $\hdots$ & $\hdots$ & $\hdots$ & $\hdots$ & $\hdots$ & $\hdots$ & $\hdots$ & $\hdots$ & $\hdots$  & $\hdots$ & $\hdots$ & $\hdots$ & $\hdots$ & $\hdots$ & $\hdots$ & $\hdots$ & $\hdots$ & $\hdots$ & $\hdots$\\ 
\noalign{\smallskip}
\hline
\end{tabular}
}
 \vskip 0.1 cm
\begin{flushleft}
    \noindent {\footnotesize{Notes: Columns are: TICv8.2 identifier, effective temperature, spectral type, TESS sectors used for the analysis, whether the UCD has flares, whether the UCD has a measured rotation period, contamination ratio, bolometric luminosity, rotational period estimated with LS, error in rotational period, rotational amplitude, error in rotational amplitude, false alarm probability, UVW galactic velocity components, UVW galactic velocity components relative to the Local Standard of Rest provided by \cite{tian2015}, errors in UVW, reference for radial velocity, galactic population membership, membership probability. \\
     All the values shown in this table were determined in this study, except for those of SpT and T$_{\rm eff}$ that were extracted from \cite{sebastian2021}. 
     $\mathrm{L_\mathrm{bol}}$ was computed only for the flaring UCDs as the flare bolometric energy divided by the flare equivalent duration. For those stars with more than one flaring event, we adopted the mean value.\\
     $^{\textit{a}}$ For those UCDs with $f_{\rm TIC}$ $>$ 0.1, the flux contamination from nearby stars may dilute the true signal amplitude. Hence, these values must be taken as lower limits.\\
     (This table is available in its entirety in machine-readable form)\\}}
\end{flushleft}
\end{threeparttable}
\end{table}
\end{landscape}

\subsection{Contamination Ratio (\textit{$f_{\rm TIC}$})} \label{contamination}

Given that the size of each TESS pixel is 21\,arcsec\,${\times}$\,21\,arcsec\,, photometric apertures used to obtain the UCDs light curves may be contaminated by the flux from nearby stars. If the UCD light curve is contaminated, any photometric variability shown might be diluted and/or, even worse, a rotation period or flare could be mistakenly attributed to a target that truly arises from a nearby star. 

The TESS Input Catalog (TICv8.2) provides an estimation that accounts for this `contamination ratio' ($f_{\rm TIC}$), as determined 
in \cite{stassun2019}. However, because only $\sim 65\%$ of the UCDs analysed in this study have a reported $f_{\rm TIC}$, we used the publicly available code \textsc{tic$\_$contam.py} \citep{paegert2021} to homogeneously calculate the contamination ratio of all the UCDs in the sample (seventh column in Table \ref{tab_1}).
Briefly, these authors identified all the point-sources with T$_{\rm mag} \lesssim 17 - 19$ at a distance within 10 TESS pixels of the target. Then, they computed their fluxes based on pre-launch PSF measurements of the field center. The size and shape of the target's aperture were defined depending on the target's TESS magnitude. Finally, $f_{\rm TIC}$ was calculated as the ratio of the flux from the objects that falls inside the aperture to the target's flux in the same aperture. 

Here, it is important to notice that the contamination ratio estimated by TESS is indicative of other stars in the field, but not robust enough to correct the measured amplitudes. Hence, we caution that the
absolute rotational amplitudes, and flare energies and amplitudes of UCDs with $f_{\rm TIC} >$ 0.1 must be taken as lower limits since these objects are 
significantly affected by the flux contamination of nearby stars.

\section{Methods}\label{methods}
\subsection{Rotation Period Measurement}\label{rotation}
Previous works \citep[e.g.][]{schmidt2015, anthony2022} revealed that most of the objects with spectral types between M4 and L4 show emission in the H$\alpha$ line, which indicates that they are magnetically active. As a direct consequence, any observed photometric variability could be interpreted as the presence of magnetic spots on the stellar surface. In this section, we present the methodology used to search for rotation periods in the UCDs of our sample.

For all the targets analysed in this work, we applied two different tools to the light curves: the Lomb--Scargle periodogram \citep[LS;][]{lomb1976, scargle1982} and the Auto Correlation Function \citep[ACF;][]{affer2012, mcquillan2013}. In the case of LS, we searched for periodic modulations between a value near the Nyquist frequency and the full observation time span. 
We adopted as the detected period, $\mathrm{P_{LS}}$, the inverse of the frequency corresponding to the highest peak in the periodogram. The period uncertainty was calculated by propagating the frequency error, which is given by the width of the peak, computed as the inverse of the baseline of the observations, i.e. the time difference between the last and first observed data point. The photometric amplitude of the rotational signal was computed as $\mathrm{\sqrt{A_{sin}^2+A_{cos}^2}}$, where $\mathrm{A_{sin}}$ and $\mathrm{A_{cos}}$ are the amplitudes of the sine and cosine terms of the best-fit model evaluated at the maximum frequency found by LS. The amplitude uncertainty was calculated through error propagation, where the uncertainties in $\mathrm{A_{sin}}$ and $\mathrm{A_{cos}}$ were estimated through the following procedure: first, we added to each flux value a random number between plus/minus its error; then, by keeping fixed the period to the value found by LS, we fitted the resulting light curve and obtained new $\mathrm{A_{sin}}$ and $\mathrm{A_{cos}}$ coefficients. We repeated these two steps 1000 times. Finally, we computed the standard deviation of the values determined for the sine and cosine amplitudes and adopted them as their uncertainties.

As a secondary verification method, we also ran the ACF to the entire light curve for each UCD. Basically, ACF assesses the degree of self-similarity of the light curve at a parameter that depends on the data cadence. It is expected to be more robust than LS in the detection of signals that change their amplitudes and phases. Once a period was detected by the ACF, we adopted as uncertainty the one calculated as in Eq.~(3) of \cite{mcquillan2013}.

In order to be confident that the measured period, after running both algorithms on the light curve, is real and not caused by instrumental systematics, it had to satisfy the following criteria: 

\begin{itemize}
\item False Alarm Probability (FAP) $\leq  0.01$\footnote{The FAP's value was computed following the method described in \cite{baluev2008}.}.
\item The standard deviation of the residuals after removing the periodic signal is smaller than or, at most, equal the standard deviation before removing the sinusoidal modulation. 
\item The period values identified by LS and ACF agree within their uncertainties.
\item The majority of the available TESS sectors for a given target shows the period value identified by LS.
\item The variability is clearly visible in the phase folded light curve. 
\item The value of the period found differs from the duration of a sector, the duration of half a sector, and the duration of the full light curve.
\end{itemize}

\noindent For further analysis, in those cases when more than one period was clearly visible in the light curve, we only considered the period with the smallest FAP value.

\subsection{Flare Identification}\label{flarefind}

We used the automated open-source code \textsc{AltaiPony} \citep{davenport2016, ilin2021} to search for flares on the TESS light curves of the 208 targets in our sample. \textsc{AltaiPony} identifies flare candidates as those with no less than three consecutive data points that positively deviate at least 3 sigma above the local scatter of the light curve and that, also, follow the criteria defined in \cite{chang2015}. For the detected candidates, the code provides as output: the times of start and end of each event, the amplitude or peak relative to the quiescent stellar flux, the equivalent duration (ED) that represents the time that would take the object to emit, in quiescent state, the same energy released during flaring state, the uncertainty in ED, and the flare's duration as the difference between the end and start times.

For the UCDs in our sample with both 2-minute and 20-second cadence light curves, we used the shortest cadence available to identify flaring events. These short-cadence data allow a better sampling and, hence, a more realistic description of the events, in particular for those with a very fast impulsive phase. Before searching for flares, we flattened the light curves. For those objects with detected rotational modulation (see Section \ref{rotation}), we subtracted the best-fit model evaluated at the maximum frequency found by LS from the light curve.  For all targets, we applied a Savitzky--Golay filter to remove any remaining uncorrected systematics in the PDCSAP light curves. 
Then, after running \textsc{AltaiPony}, we inspected by eye all of the events identified by the code and kept those with the typical flare profile (i.e. a suddenly increase and exponential decay in flux) or a multi-flare shape \citep[e.g.][]{gunther2020}. 

To measure the bolometric energy of the detected flares, we followed the work of \cite{howard2022} and used the equation:

\begin{equation}
      \mathrm{E_{\mathrm{bol}}} = \frac{\mathrm{ED} \times \mathrm{L_{\mathrm{TESS}}}}{c}.
      \label{ec3}
\end{equation}

\noindent Here, $\mathrm{L_{\mathrm{TESS}}}$ is the quiescent luminosity considering the TESS CCD response. This quantity was computed for each target through the luminosity-flux-distance relationship by adopting the flux of a star with TESS magnitude of zero derived by \cite{sullivan2015}, T$_{\rm mag}$ from the TICv8.2, and the distance from the \Gaia-DR3 catalog \citep{gaia2016, gaia2022, katz2022}. When the distance was not available in the \Gaia-DR3 catalog, we extracted it from the TICv8.2.
The constant `c' is the correction factor for the TESS CCD response calculated by \cite{howardmcgregor2022} assuming a flare with a continuum component characterized by a 9000 K blackbody. This constant has a value of c = 0.19 representing the energy fraction released in the TESS band during the flare.

\subsection{Planetary Transit Search}\label{TransitSearch}

According to the core accretion theory \citep{pollack1996}, the small size and low-mass protoplanetary disks around late-type stars \citep{andrews2013, pascucci2016} would create a favorable environment for the formation of small rocky planets around these objects \citep{raymond2007, alibert2017}. Additionally, theoretical studies about planetary formation \citep[e.g.][]{mulders2015}, point out that the occurrence rate of terrestrial planets is higher for M stars compared to FGK stars. Hence, late-type objects are ideal candidates to host close-in, Earth-like planets.
Furthermore, small planets orbiting UCDs produce deeper transits and larger RV semi-amplitudes than small planets around solar-type stars. Due to their low luminosities, the habitable zone is closer to the UCD host than FGK hosts, increasing the chances of detecting planets orbiting within \citep{irwin2009}. 

Albeit these advantages, UCD planetary systems remained elusive for a long time due to their emission being predominantly at IR wavelengths. However, in the last decade, different projects emerged with the purpose of searching for planets around UCDs using ground based facilities, such as SAINT-EX \citep{GomezMaqueo2023}, SPECULOOS \citep{delrez2018}, EXTRA \citep{shaklan2015}, PINES \citep{tamburo2022}, CARMENES \citep{quirrenbach2018}, SPIRou \citep{donati2018}. To date, a few planetary systems around UCDs were already confirmed \citep{gillon2016, gillon2017, anglada-escude2016, zechmeister2019}.

We searched for signs of planetary transits in the 2-minute cadence TESS light curves of the 208 targets of our sample. For those UCDs with detected rotational modulation, we first subtracted the best-fit model at the rotational period value from the PDCSAP light curve. Then, we executed a time-windowed slider algorithm based on Tukey's biweight provided by the open-source package \textsc{w{ö}tan} \citep{hippke2019a} on the time-series of all the targets to remove any remaining systematics. The search for transit signals was carried out with the Transit Least Squares (\textsc{TLS}) algorithm \citep{hippke2019b}, giving it as input the detrendend light curve and the quadratic limb-darkening coefficients of each target extracted from the TICv8.2. To validate the signal detected by the algorithm and mark the object as `planet candidate', it had to satisfy the following conditions:

\begin{itemize}
    \item Signal Detection Efficiency (SDE) $> 6.0$.
    \item More than one transit detected in the light curve.
    \item Transit clearly visible in the phase folded light curve and the best-model found by \textsc{TLS} well-fitted to the data.
    \item Agreement between the measured depths of the detected transits, within errors. 
\end{itemize}

\noindent None of the initially detected signals fulfilled all of these criteria. Only a few UCDs present signals that satisfy at least one of the criteria. 
However, after a more rigorous inspection and a reanalysis of the data, they were finally excluded as spurious. Hence, we did not identify any possible transiting planet (or eclipsing stellar companion) candidate orbiting around any UCD in our sample.

\section{Results}\label{results}
\subsection{Measured Rotation Periods and Amplitudes}

\begin{figure*}
    \centering
     \includegraphics[width=1.1\textwidth,trim=1.0cm 0.2cm 0cm 0.7cm, clip]{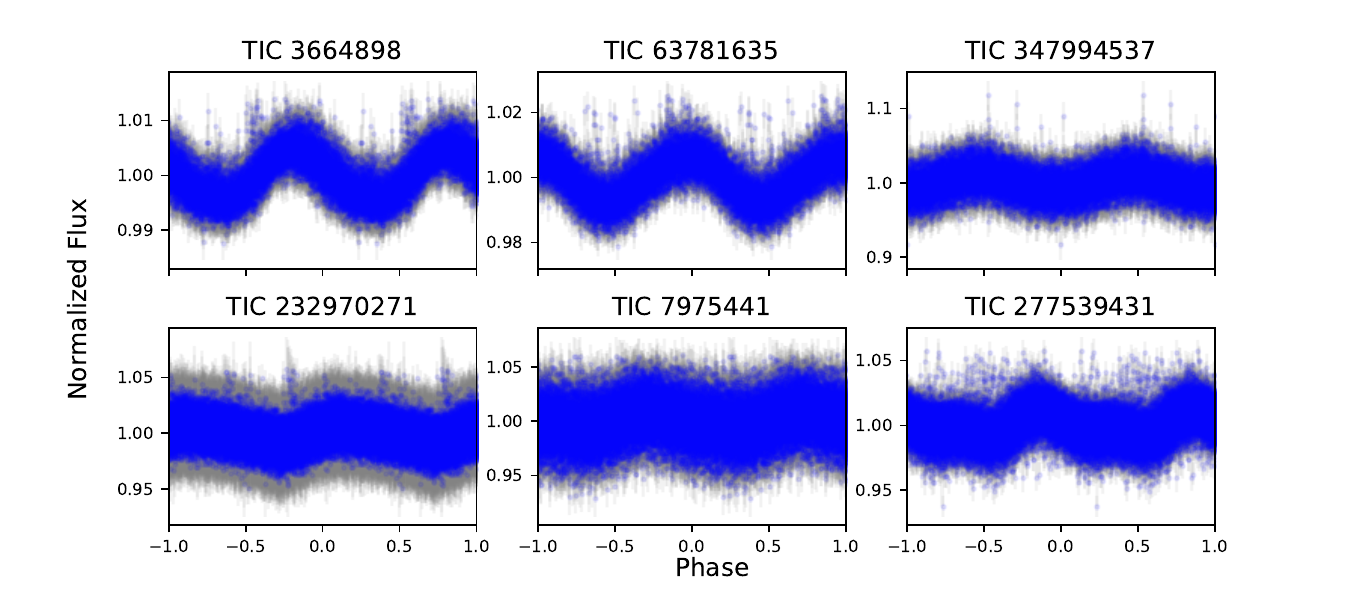}
   \caption{Phase-folded light curves of six UCDs showing the measured rotational modulation. Two full cycles are shown each panel, repeating the data twice. The 6 UCDs have been chosen to showcase the best- (top row) and worse-case (bottom row) scenario in the detection of rotational modulation. The y-axis scale of the top row changes from panel to panel. 
   The measured periods are: $0.458 \pm 0.038$\,d (TIC~3664898), $0.276 \pm 0.043$\,d (TIC~63781635), $0.986 \pm 0.040$\,d (TIC~347994537), $0.511 \pm 0.004$\,d (TIC~232970271), $0.464 \pm 0.001$\,d (TIC~7975441), and $0.190 \pm 0.001$\,d (TIC~277539431).
   }
              \label{LCprot}
\end{figure*}

We found that 87 UCDs in our sample fulfill the criteria described in Section \ref{rotation}, indicating that $\sim 42\%$ present a measurable rotational modulation in their TESS light curves. The measured periods span from 2.02 hours to 4.63 days, while their amplitudes
range from  0.0009 to 0.1986\,mag, as measured directly from the TESS light curves, in agreement with previous findings \citep{seli2021, medina2022, milespaez2023}. In Table \ref{tab_1}, we present the rotational periods and absolute amplitudes measured in this work, and Fig. \ref{LCprot} displays selected light curves with detected periodic photometric variability. The six UCDs shown in the figure have been selected such that the top three panels represent clearly sinusoidal rotational modulation and the bottom three show the lower-limit of rotational modulation, including non-sinusoidal modulation patterns. 
Figure~\ref{stats_all} shows the distribution of objects as a function of spectral type, as well as the distribution of those with measured rotation period (pink and orange colours).

\subsection{Active vs non-active UCDs}
We grouped the targets analysed in this work into two categories: `active' and `non-active'. In the first one, we included objects for which we were able to detect rotational modulation or flares or both, and in the second category, those with neither rotational modulation nor flares detected.  Out of 208 objects in our sample, we found that the `active' UCDs are:

\begin{itemize}
    \item 31 ($\sim$ 15 $\%$) only show rotational modulation,
    \item 47 ($\sim$ 23 $\%$) only present at least one flare,
    \item 56 ($\sim$ 27 $\%$) have a detected rotational modulation and flares.
\end{itemize}

\begin{figure*}
    \centering
     \includegraphics[width=8.5cm]{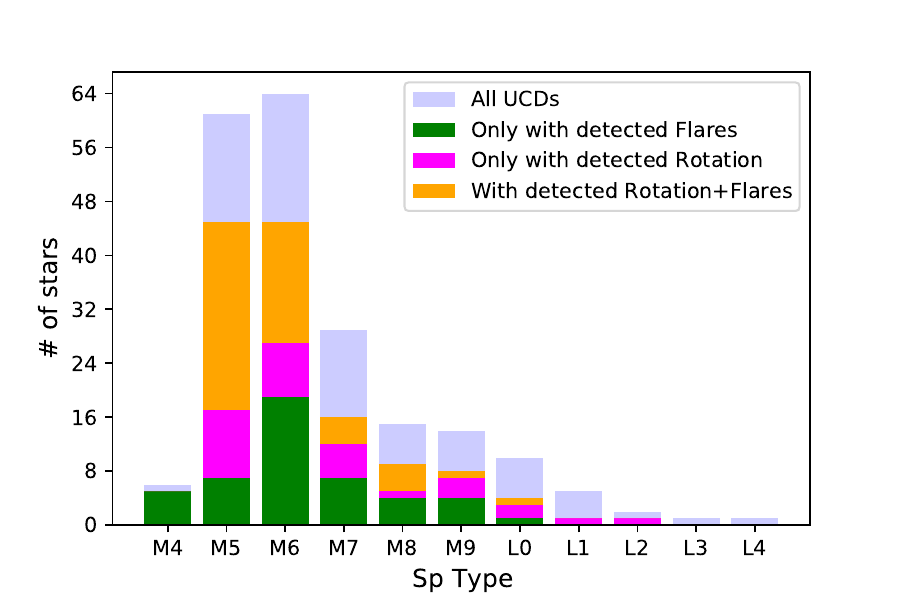}
     \includegraphics[width=8.5cm]{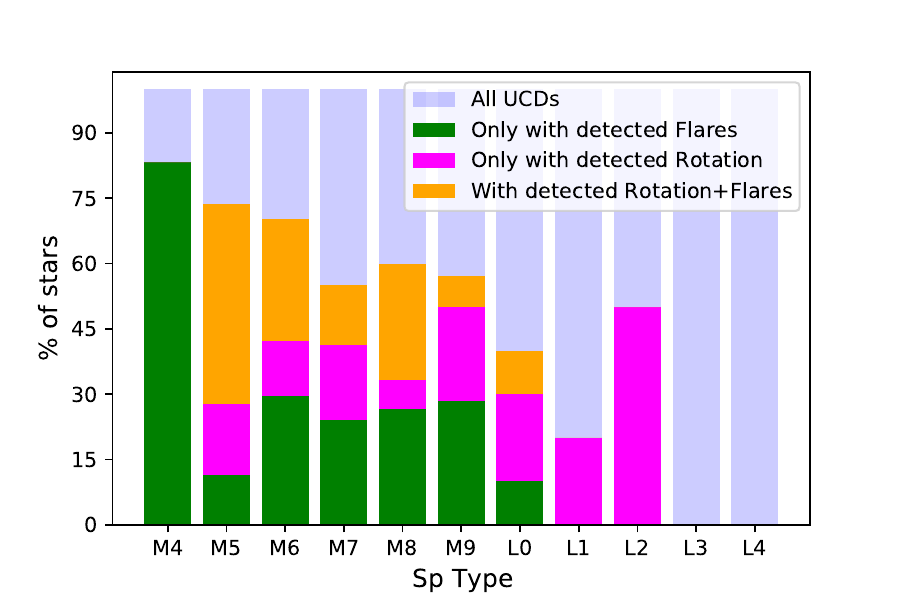}
   \caption{Bar graph showing the distribution of UCDs in our sample as a function of spectral type. Each color-coded bar represent a specific group. {\it Left:} Total number of UCDs in the sample per spectral-type bin (light blue).  In green, UCDs with at least one detected flare in their TESS light curve. In pink, UCDs with measured rotation periods. In orange, UCDs with both detected flares and measured rotation periods. {\it Right:} Percentage of UCDs per spectral type in the full sample (light blue), only with detected flares (green), only with measured rotation (pink) and with both measured rotation and flares (orange). 
   For our sample, earlier spectral types (M4--M6) tend to show more signatures of activity than later type objects (M7--L4). 
   }
              \label{stats_all}
\end{figure*}

\noindent This means that $\sim 64\%$ of the UCDs in our sample (i.e. 134 objects) are `active' targets. In Figure \ref{stats_all}, we show histograms of the number (left) and fraction (right) of objects per spectral type and highlight the UCDs that present some signature of activity. In general terms, the figure shows that earlier spectral type targets (M4--M6) tend to be more active than later-type ones (M7--L4).

\begin{figure*}
    \centering
     \includegraphics[width=1.1\textwidth,trim=1.5cm 0.25cm 0.25cm 1.2cm, clip]{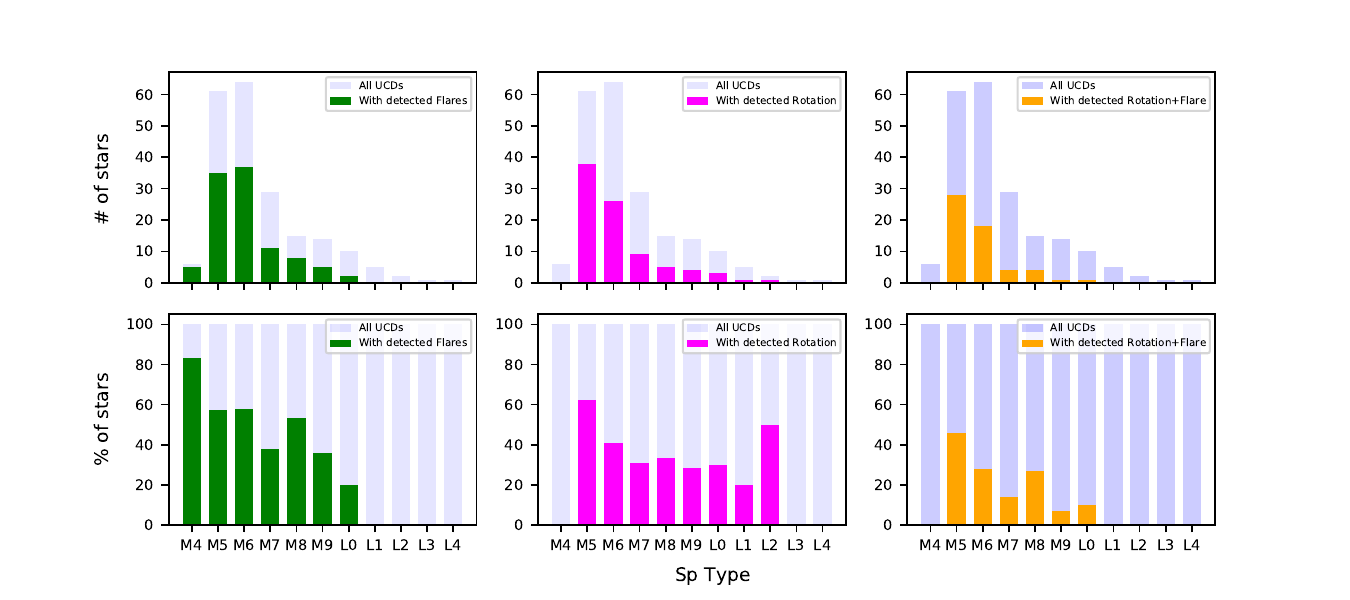}
   \caption{Histograms of the number (top in light blue) and fraction (bottom in color) per spectral type of flaring UCDs with and without detected rotational modulation (left, in green), UCDs with detected rotational modulation with and without flares (middle, in pink), and UCDs with detected rotational modulation and flares (right, in orange). In the left panel, we note a peak in the number and fraction of flaring around early-type M4--M6 UCDs that decreases toward late spectral types (M7--L4). We find the same result in the mid and right panels, where the number and fraction of objects with detected rotation periods is maximum around M4--M6.}
              \label{stats_6fig}
\end{figure*}

Figure \ref{stats_6fig} shows in separate panels the proportion of UCDs per spectral type that have at least one identified flare (green), measured rotation period (pink), or both (orange), compared to the full UCD sample. Regarding the flaring activity (left panel), we found that earlier type objects (M4--M6) in our sample show a peak in the number of targets with detected flares with $\sim$60 to 80\% UCDs in those spectral bins. The number of UCDs with detected flaring activity
decreases toward later spectral types (M7--L4), with no flares detected for spectral types L1 through L4 in our sample.  This is in agreement with previous photometric studies. For example, \cite{medina2022} found that the fraction of stars with flares is maximum at $0.15-0.2\,M_{\odot}$ ($\sim$\, M4--M5) and decreases for later spectral types. \cite{yang2023} also observed an increasing trend in flaring from M0 to M5 type stars and a posterior downward trend from M5 to M7 through the analysis of TESS data. Additionally, the works of \cite{yang2017}, \cite{rodriguez2020}, \cite{gunther2020} used independent photometric data (\textit{Kepler}, ASAS, and TESS, respectively) and found that the fraction of stars with confirmed flares peaks at spectral type M4--M5. On the other hand, the results from \cite{murray} show an increasing number of UCDs with flaring from M4 to M7 spectral types followed by a decline toward L0, with no flares detected in L1--L2 objects. Considering objects with detected rotation periods (middle and right panels), we found the same behaviour as for those that only flares, where the fraction of targets showing activity signatures is maximum around M4--M6, which also agrees with previous results \citep{gunther2020}. 

Concerning the statistics, our results indicate that 42$\%$ of the UCDs in the sample have a measured rotation period, in agreement with \cite{newton2016}. These authors used ground-based observations from the MEarth Project \citep{nutzman2008} of 387 nearby, mid-to-late M dwarfs ($\sim$\, M3 and later) and measured rotational modulation on 47$\%$ of the targets. Meanwhile, \cite{seli2021} analysed 30-minute cadence TESS data of 248 TRAPPIST–1-like ultra-cool dwarfs (i.e. objects closer than 0.5 magnitudes to TRAPPIST–1 on the \Gaia\ color-magnitude diagram at a distance up to 50\,pc away) and found that only 17$\%$ (42/248) present periodic light curve modulation with 21 of these UCDs also showing flares. This could be a consequence of the relatively longer cadence of the TESS observations. On the other hand, the analysis by \cite{mcquillan2013} of more than 2400 main-sequence M stars observed by \textit{Kepler} \citep{borucki2010} revealed that 63.2$\%$ of the objects in their sample have detected rotation periods ranging from 0.37 to $\sim 70$\,days. \cite{raetz2020} found rotation periods as long as 80 days for about 82$\%$ of all targets in their K7--M6 sample (56 objects in total) using K2 long- and short-cadence data. Similar statistics was found by \cite{raetz2019} through the analysis of 430 K8--M7 stars considering only K2 long-cadence data. In comparison, the lower percentage of UCDs with detected rotational modulation found in this study, might be a consequence of the time span ($\sim 27$~d for each TESS sector) of the analysed observations which makes finding periodicities larger than 5 days more challenging. This would imply that we might be biased against detecting modulation on those objects with longer-term photometric variability.

Additionally, we notice that the non uniformity in the number of the available TESS sectors in which each of the UCD analysed in this study was observed (from only one to 26 sectors in some objects) might introduce a bias in the detection of flare events and rotation periods towards UCDs with high flare rates and fast rotation.

\subsection{Searching for correlations between parameters of active UCDs}

\subsubsection{No correlation between rotation parameters and stellar properties}

\begin{figure*}
    \centering
     \includegraphics[width=8cm]{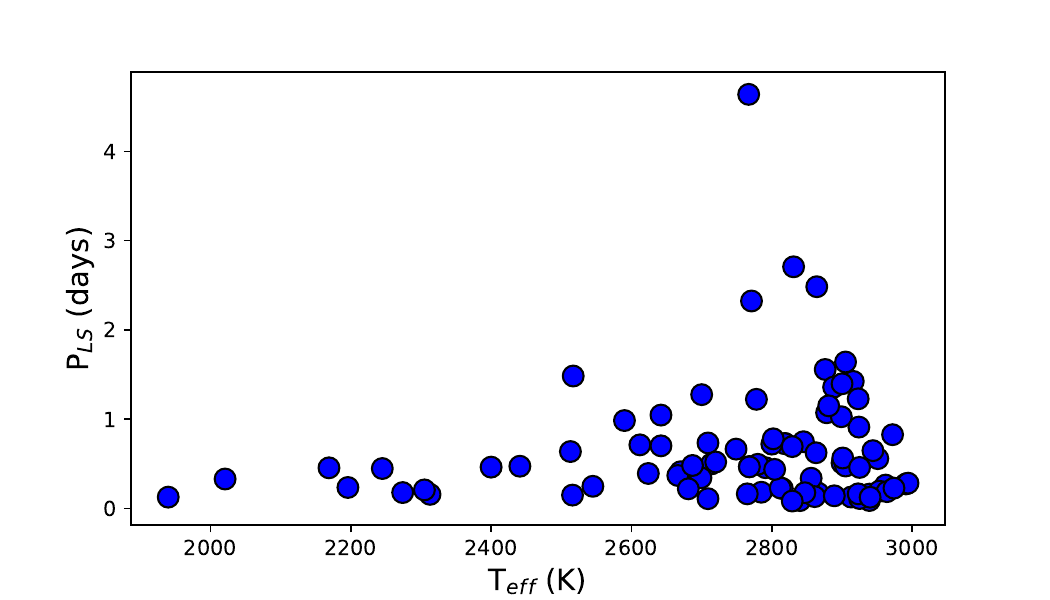}
     \includegraphics[width=8cm]{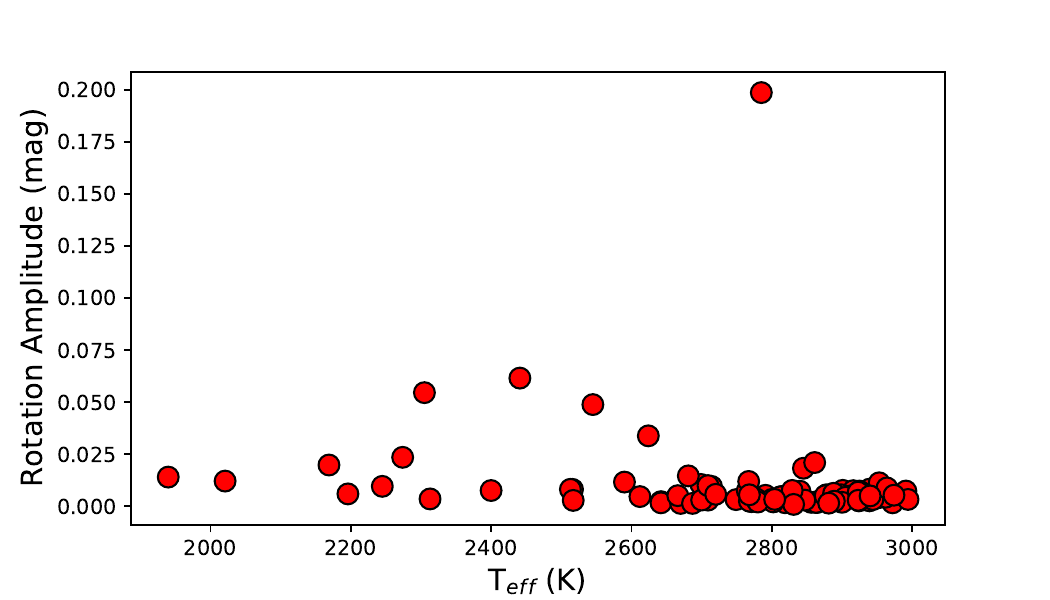}
   \caption{\textit{Left:} Effective temperature versus rotation period for the 87 targets with measurable rotation in the sample. No clear correlation is observed between these parameters. We note that the high dispersion only observed in the periods between 2700 $<$ $\mathrm{T_{eff}}$ $<$ 3000 $\mathrm{K}$, may be a consequence of the small number of targets with $\mathrm{T_{eff}}$ $<$ 2700 $\mathrm{K}$ that have a measurable rotation in our sample. \textit{Right:} Effective temperature versus rotation amplitude for the targets with measurable rotation shown in the left panel. 
   Also in this case, there is no evident correlation between the measured rotational amplitude and the effective temperature. The high dispersion found in the values of rotational amplitude in the 2200 $\leq$ $\mathrm{T_{eff}}$ $\leq$ 2600 $\mathrm{K}$ range may be an effect of the small number of targets in this group (only 36) compared to those in early-spectral types (60 and 64 targets, respectively).
   }
              \label{TeffvsProtAmp}
\end{figure*}

We investigated possible trends between the parameters that characterize the rotational modulation (i.e. period and amplitude) and the effective temperature and spectral type of the UCDs in our sample. In the left panel of Fig.~\ref{TeffvsProtAmp}, we show a plot of effective temperature ($\mathrm{T_{eff}}$) from \citet{sebastian2021} as a function of our measured period, $\mathrm{P_{LS}}$, for the 87 objects with detected rotational modulation. We note a higher dispersion in the measured periods found for   effective temperatures of $2700 < \mathrm{T_{eff}} < 3000\,\mathrm{K}$, compared to the rest of the sample. However, caution must be taken due to the scarce number of targets with $\mathrm{T_{eff}} < 2700\,\mathrm{K}$ that have a measurable rotation in our sample. No clear trend is revealed when the rotational periods are analysed as a function of spectral type. A lack of correlation is also found if targets are separated in those with detected flares and without detected flares. Objects with spectral type M4, L3 and L4 are not shown because no rotational modulation was detected in the TESS light curves that were analysed. 
Additionally, in the right panel of Fig. \ref{TeffvsProtAmp}, we show a plot of $\mathrm{T_{eff}}$ as a function of rotation amplitude for the same objects presented in the left panel.
No correlation is observed between the amplitude of the rotational modulation and the effective temperature or spectral type. 
Although UCDs in the 2200 $\leq$ $\mathrm{T_{eff}}$ $\leq$ 2600 $\mathrm{K}$ range, i.e. spectral types 
M7.5 to L0,  
present a higher dispersion in the values of rotation amplitude, this seems to be an effect of the small number of targets in this group (only 36) compared to those with spectral types M5 and M6 (60 and 64 targets, respectively).
All these results point toward a rotational modulation in the light curves that might depend on other factors, besides the difference in the energy transfer mechanism in fully and partially-convective objects.

\subsubsection{No correlation between flare parameters and spectral type}

\begin{table*}
\centering 
\begin{threeparttable}
\caption{Flares' main parameters determined in this work.}
\label{tab_2}  
\begin{tabular}{ccccccc}
\hline 
\noalign{\smallskip} 
 \multirow{2}{3em}{TIC ID} & t$_{\rm start}$ & t$_{\rm end}$ & ED$^{\textit{a}}$ & Amplitude$^{\textit{a}}$ & E$_{\rm bol}$ &  Cadence \\ 
 & (TBJD) & (TBJD) & (s) & (relative flux) & ($\mathrm{erg\,}$) & (s) \\
\noalign{\smallskip} 
\hline 
\noalign{\smallskip} 
420130591 & 2635.791 & 2635.803 & 45.045 & 0.168 & $2.310 \times 10^{32}$ & 20 \\
420130591 & 2394.111 & 2394.119 & 44.977 & 0.201 & $2.306 \times 10^{32}$ & 20 \\
420130591 & 1820.404 & 1820.411 & 27.049 & 0.119 & $1.387 \times 10^{32}$ & 120 \\
420130591 & 2732.656 & 2732.669 & 23.858 & 0.068 & $1.223 \times 10^{32}$ & 20 \\
420130591 & 2437.652 & 2437.662 & 23.479 & 0.067 & $1.204 \times 10^{32}$ & 20 \\
420130591 & 1827.117 & 1827.124 & 23.419 & 0.069 & $1.201 \times 10^{32}$ & 120 \\
420130591 & 2414.168 & 2414.178 & 20.355 & 0.046 & $1.044 \times 10^{32}$ & 20 \\
420130591 & 2443.971 & 2443.978 & 19.884 & 0.068 & $1.020 \times 10^{32}$ & 20 \\
420130591 & 1845.360 & 1845.367 & 18.053 & 0.033 & $9.257 \times 10^{31}$ & 120 \\
420130591 & 1840.594 & 1840.601 & 17.048 & 0.049 & $8.742 \times 10^{31}$ & 120 \\
420130591 & 2396.179 & 2396.187 & 16.540 & 0.052 & $8.481 \times 10^{31}$ & 20 \\
420130591 & 1805.529 & 1805.536 & 16.038 & 0.040 & $8.224 \times 10^{31}$ & 120 \\
420130591 & 2752.709 & 2752.715 & 14.530 & 0.056 & $7.451 \times 10^{31}$ & 20 \\
420130591 & 2587.945 & 2587.949 & 13.903 & 0.073 & $7.129 \times 10^{31}$ & 20 \\
420130591 & 2618.558 & 2618.562 & 13.668 & 0.072 & $7.008 \times 10^{31}$ & 20 \\
$\hdots$  & $\hdots$ & $\hdots$ & $\hdots$  & $\hdots$ & $\hdots$ & $\hdots$ \\
\noalign{\smallskip}
\hline
\end{tabular}
\vskip 0.05 cm
\noindent {\footnotesize{Notes: Columns are: TICv8.2 identifier, flare start time, flare end time, equivalent duration, flare amplitude, bolometric energy, TESS data cadence.\\
$^{\textit{a}}$ For those UCDs with $f_{\rm TIC}$ $>$ 0.1, the flux contamination from nearby stars may dilute the true flare ED/amplitude. Hence, these values must be taken as lower limits.\\
(This table is available in its entirety in machine-readable form)\\}}
\end{threeparttable}
\end{table*}

\begin{figure*}
    \centering
     \includegraphics[width=1.1\textwidth,trim=1.0cm 0cm 0cm 1.2cm, clip]{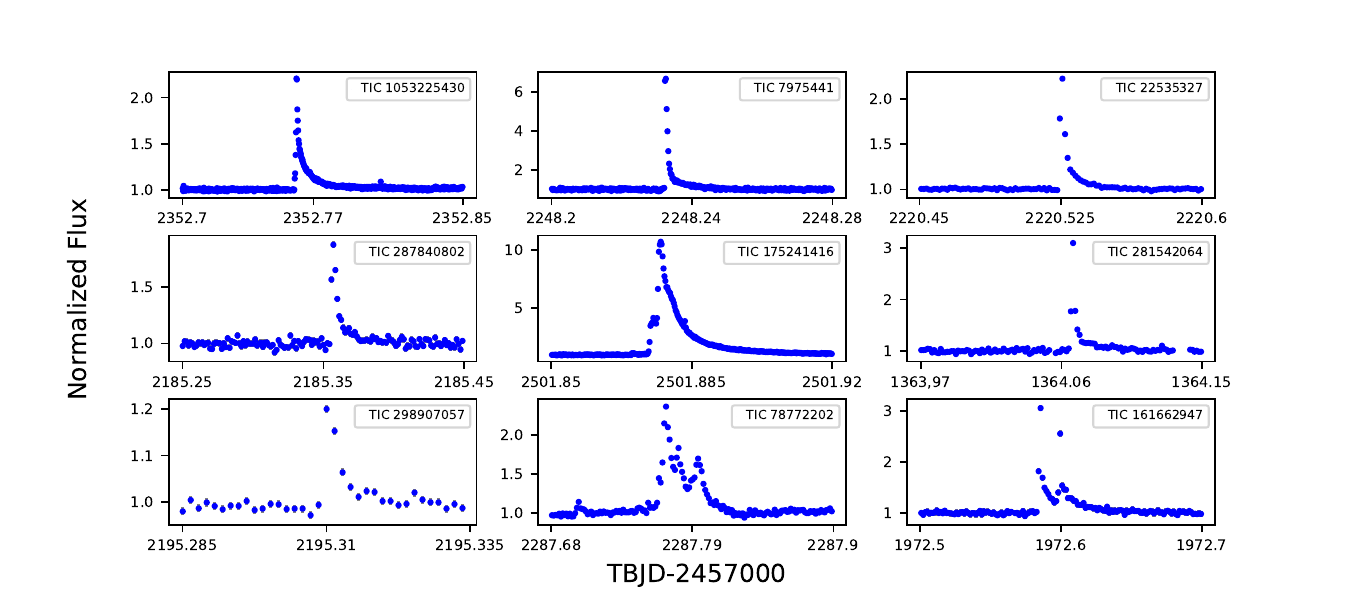}
   \caption{Nine flaring events corresponding to nine independent UCDs analysed in this work. These flares have been chosen to showcase the different kind of flaring events that were identified in the analysed TESS light curves. The top row shows well-sampled flare events with typical profiles. The middle row presents noisy flares, and the bottom row displays a poorly-sampled noisy event on the left panel and multi-flare shape events on the mid and right panels. Error bars are shown, but in most cases, they are smaller than the symbol's size. Both axes in each panel have been optimized to each event. }
              \label{flares}
\end{figure*}

Following the prescription indicated in section \ref{flarefind}, we found a total of 778 flares in the TESS light curves of 103 objects, which represents $\sim 49.5\%$ of the total sample. In Table \ref{tab_2}, we present the main parameters of the 778 identified flares and, in Fig. \ref{flares}, selected flaring events are shown. 

\begin{figure*}
    \centering
     \includegraphics[width=8.5cm]{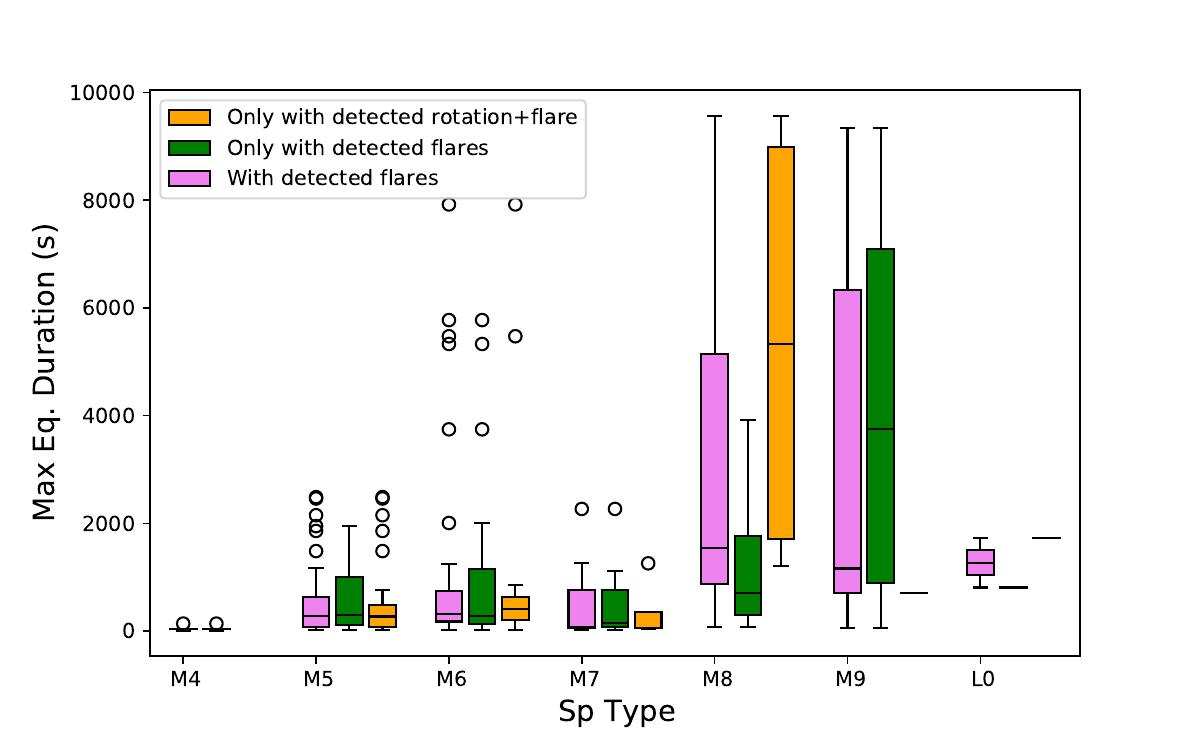}
     \includegraphics[width=8.5cm]{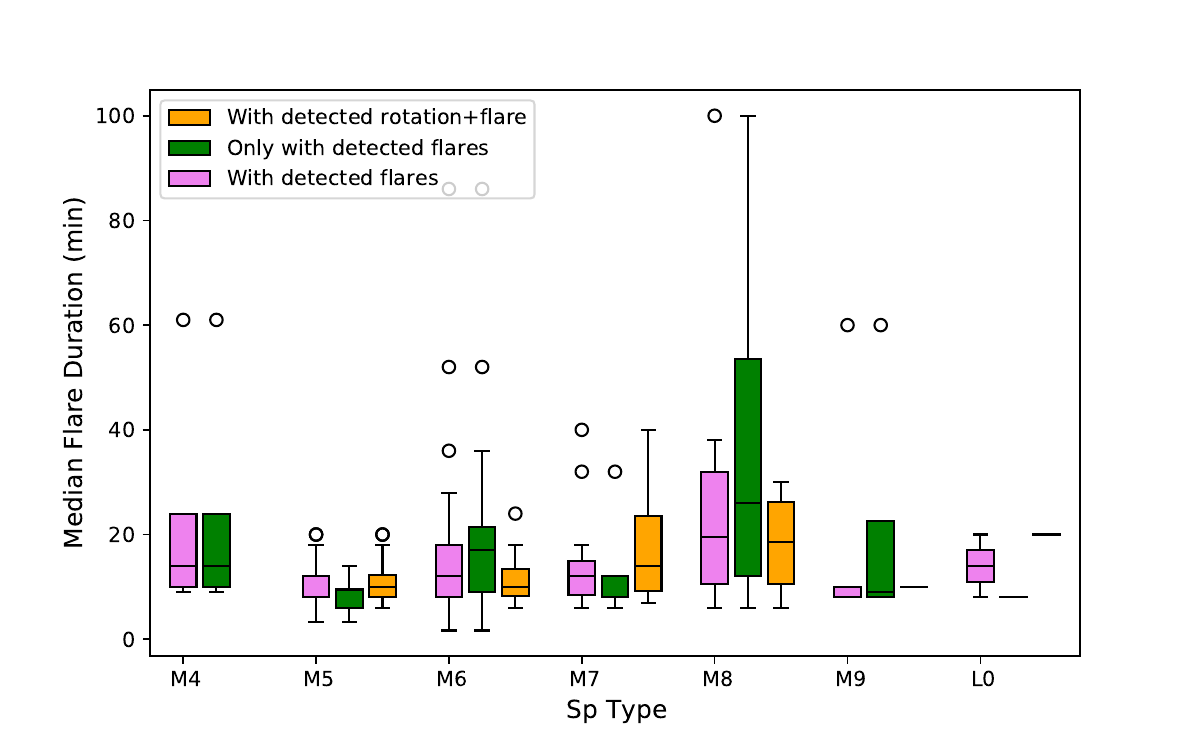}
   \caption{{\it Left:} Box plot of the maximum equivalent duration per spectral type for the 103 targets with detected flares.
   For the UCDs with more than one detected flare, we chose the longest equivalent duration event. Although median values agree within the interquartile range, we note a slight increase from M7 to M8--L0 stars, where a peak is reached. Nonetheless, it is worth noting that this trend can be due to the small number of M8--L0 UCDs (15 in total) compared with the 88 stars with M4--M7 spectral types.
   {\it Right}: Box plot of the flare median duration per spectral type for the 103 targets with detected flares. The median duration was estimated from the duration of all flares detected per object. No evident correlation is observed, and median values agree within the interquartile range. We note that events in M8 UCDs seem to last slightly longer than the rest. However, this may be a consequence of the small number of targets (only 8) that constitute this group. Here, targets that flare and have a detected rotation period are indicated in orange color, those that only flare in green, and in pink the summed targets of the other two.  
   }
              \label{MaxEDvsSpT}
\end{figure*}

We explored possible correlations between the parameters associated with flares (i.e. equivalent duration, flare amplitude, duration, and flare rate) and spectral type.
In particular, we investigated if, compared with early-type UCDs, late-type objects present more energetic and longer lasting flaring events. In the left panel of Fig.~\ref{MaxEDvsSpT}, we present a box plot of the maximum equivalent duration per spectral type for the 103 targets with detected flares.
For UCDs with more than one detected flare, the longest equivalent duration event was chosen. 
Objects with spectral types L1--L4 are not shown because none of them present flaring events.
We can see that even though median values (marked with an horizontal black line inside each box) agree within the interquartile range, these seem to slightly increase from M7 to M8--L0 stars, where a peak is reached. A similar behaviour is observed for the median equivalent duration per spectral type. Here, the values of the median equivalent duration were computed considering all flares detected per object. Nonetheless, these results must be taken with caution because only 15 UCDs are M8--L0 compared with the 88 that have a spectral type between M4 and M7.
\cite{murray} also found a peak, but shifted at M7 targets and concluded that, in comparison, spectral types later than M7 show a real absence of high energetic events.
However, in contrast with the study of \cite{murray}, we do observe the same behaviour for the maximum and median flare amplitude per spectral type, which is consistent with the positive correlation between equivalent duration (energy) and amplitude of flares found in previous studies (see Section \ref{flarepar}).
Additionally, in the right panel of Fig. \ref{MaxEDvsSpT}, we show a box plot of flare's median duration in minutes for the 103 targets with detected flares. This median duration was calculated from the span of all flares detected per object. No clear trend is observed and median values are in agreement within the interquartile range. Events associated with M8 UCDs seem to last slightly longer than the rest. However, given that only 8 targets constitute this group, which is a small number compared with the size of the samples for the other spectral types, this trend must be taken with caution. Also, there is no evident correlation between flare rate, measured in number of events per day, per spectral type.

\subsubsection{No correlation between flare and rotation parameters}

For the 56 UCDs with detected rotational modulation and flare events, we searched for possible correlations between the two rotation parameters (period and amplitude) and equivalent duration (maximum and median value), and flare amplitude (maximum and median value), and correlations between amplitude of the rotational modulation and flare rate, and flares median duration. No correlations between these parameters were identified. We found a similar result after plotting the value of the rotation period against flare rate and flares median duration for all the 56 targets with detected rotational modulation and flares in the sample. Previous studies \citep{newton2017, gunther2020, murray} found that very fast rotators have a higher likelihood of flaring than slow rotators. In our sample, we are unable to confirm these conclusions, given that the majority of our targets have detected rotation periods $\lesssim 2$~days (and are thus considered fast rotators), whilst the aforementioned works expand the range of rotation periods to $> 5$~days. Finally, we found no correlation between rotation amplitude and rotation period for the UCDs in our sample, which confirms previous results from \cite{newton2016}, and \cite{medina2020, medina2022}.

\subsection{Galactic Kinematics and Active UCDs}

We cross-matched the coordinates from the TESS Input Catalog with the \Gaia-DR3 catalog adopting a search radius of 15 arcsec. We considered this large search radius because the UCDs in our sample have high proper motions. When more than one object fell in the search area, we chose the one with the largest $\mathrm{G_{RP}}-\mathrm{G_{BP}}$ color (i.e. the reddest one) and extracted its proper motion, parallax and radial velocity values. For those targets with no \Gaia\ information, we searched catalogs available in the SIMBAD database for each object, that included the parameters mentioned above. Then, we computed the galactic velocity components $\mathrm{U_{LSR}}$, $\mathrm{V_{LSR}}$, $\mathrm{W_{LSR}}$ with their errors for all the UCDs in our sample, following the methodology described in \cite{johnson1987}. We adopted the solar velocity components relative to the Local Standard of Rest provided by \cite{tian2015}, ($\mathrm{U_e}$, $\mathrm{V_e}$, $\mathrm{W_e}) = (9.58, 10.52, 7.01)\,\mathrm{km\, s^{-1}}$. Finally, we used the criteria employed in \cite{reddy2006} to determine the membership probability of each UCD to the thin disk, thick disk or halo, including transition regions between the three Galactic populations. 

\begin{figure*}
   \centering
   \includegraphics[width=1.05\textwidth,trim=1.2cm 0.2cm 1cm 1cm, clip]{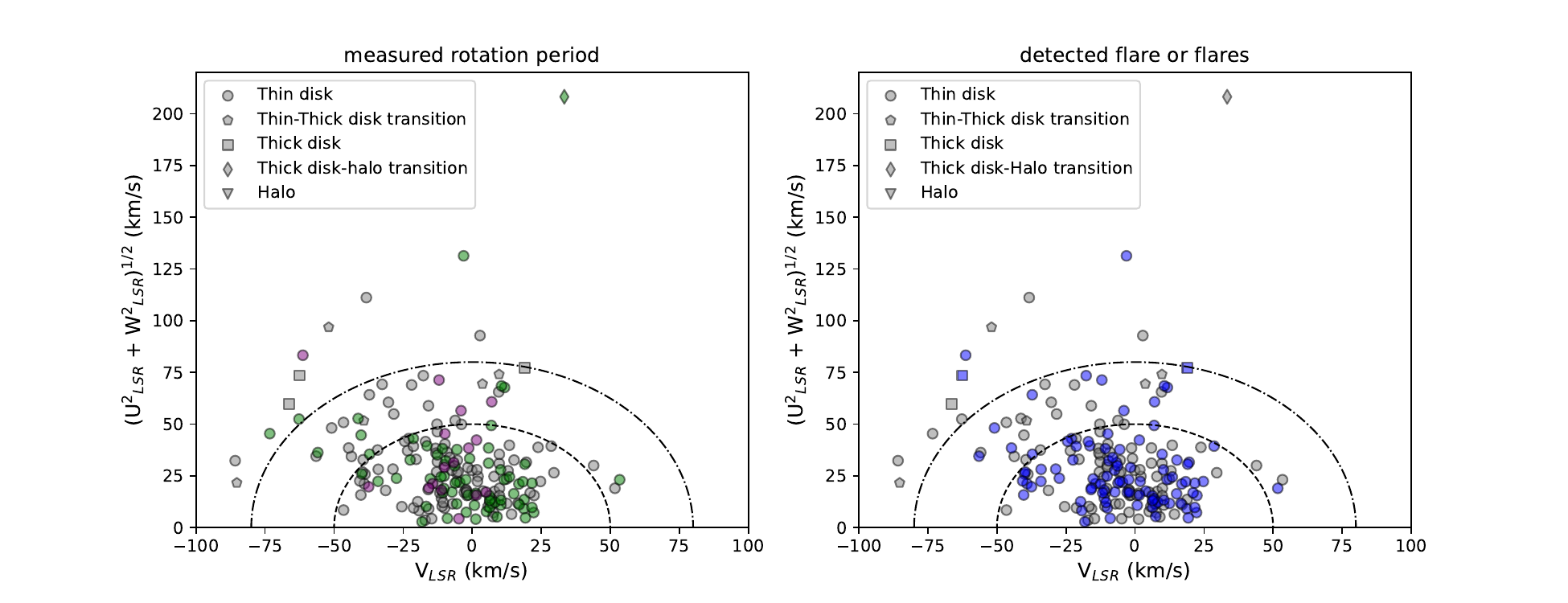}
   \caption{Toomre diagrams of 196 UCDs in our sample. 
   Each galactic population is shown with a different symbol (see legend).
   Dashed and dashed-dot lines represent velocity contours of 50 and $80\,\mathrm{km\,s^{-1}}$, respectively. For clarity, the target that belongs to the halo is not shown. \textit{Left panel:} Green, violet, and gray colors indicate UCDs with measured rotation periods $\leq 1$\,d, measured rotation periods $> 1$\,d, and without a measured rotation period, respectively. \textit{Right panel:} Blue and gray colors mark UCDs with at least one detected flare and without identified flares, respectively. Both panels show that active UCDs are spatially distributed around almost all Galactic components.}            \label{uvw}
\end{figure*}

\begin{figure}
   \centering
   \includegraphics[width=9cm]{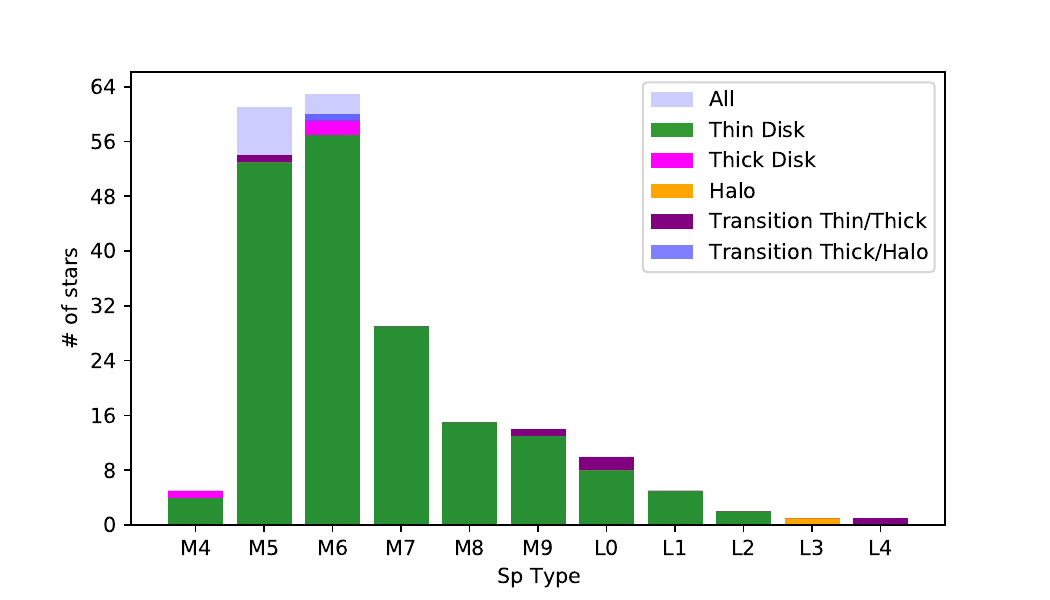}
   \caption{Bar graph of the number of objects per spectral type that belong to the thin disk, thick disk, halo, and transition zones.}
        \label{uvwh}
\end{figure}

We measured the $\mathrm{U_{LSR}}$, $\mathrm{V_{LSR}}$, $\mathrm{W_{LSR}}$ values for 196 UCDs in our sample, of which we have: 186 ($\sim 94.9\%$) from the thin disk, three ($\sim 1.5\%$) from the thick disk, one ($\sim 0.5\%$) from the halo, five ($\sim 2.5\%$) that belong to the transition zone between the thin and the thick disk, and one ($\sim 0.5\%$) from the transition zone between the thick disk and the halo. We were not be able to measure $\mathrm{U_{LSR}}$, $\mathrm{V_{LSR}}$, $\mathrm{W_{LSR}}$ for 12 objects, given the lack of radial velocity measurements in the literature. These results are presented in Table \ref{tab_1}. 
In Fig. \ref{uvw}, we show the location of the UCDs in a Toomre diagram and their galactic membership as a function of spectral type in Fig. \ref{uvwh}. In Fig. \ref{uvw}, different symbols mark targets from different galactic components. The target that belongs to the halo is an L3.1V UCD and it is not shown for a better visualization of these plots, as it is located at ($\sim 753 \mathrm{km\, s^{-1}}$, $\sim 638 \mathrm{km\, s^{-1}}$). The halo UCD did not have a measured rotational modulation nor detected flares. In the left panel, we distinguish UCDs with measured rotation periods, $\mathrm{P_{LS}}$ $\leq$ 1 d, and $\mathrm{P_{LS}}$ $>$ 1 d in green and purple, respectively, whilst those objects without a measured rotation period are marked in gray.
It can be seen that UCDs with measured rotational modulation are found in almost all of the galactic populations, including the transition zones. A similar result is observed in the right panel. UCDs with and without detected flares (blue and gray coloured symbols, respectively) are spatially distributed around almost all Galactic components.

\subsection{Superflares}\label{superflare}

According to \cite{schaefer}, superflares are defined as flares with bolometric energies from $10^{33} \mathrm{erg\,}$ to as high as $10^{38} \mathrm{erg\,}$. Following this definition, in the present study, we identified 56 superflares from 33 stars (27 M5, 21 M6, 2 M7, 4 M8, and 2 M9) with $\mathrm{E_{bol}}$ between $1.0 \times 10^{33}$ and $1.1 \times 10^{34} \mathrm{erg}$.
As a comparison, \cite{howardmcgregor2022} explored the time-resolved properties of flares in a sample of 226 M stars using TESS 20-second cadence mode data and discovered 428 superflares, with 27 events showing energies $> 10^{35} \mathrm{erg\,}$. Additionally, \cite{raetz2020} found 91 superflares on 46 rotating M dwarfs observed with K2, while \cite{murray} did not detect superflares in their sample of UCDs. In our work in particular, the most energetic flare of the entire sample, which released a bolometric energy of 1.15 $\times$ $10^{34}$ $\mathrm{erg\,}$ over 1.66 hours, occurred on the UCD TIC 175241416, an M6 star in the North Hemisphere with a T$_{\rm mag}$ $=$ 13.326, observed in three TESS sectors.

\subsection{Identifying correlations between flare energy, amplitude and duration}\label{flarepar}

Following previous studies \citep{hawley2014, silverberg2016, yang2023}, we searched for correlations between flares' bolometric energy, amplitude, and duration for 102 flaring objects. In this analysis, we excluded the target TIC 318801864, an M9 UCD for which we were unable to estimate the bolometric energy of its unique flare event given that neither \Gaia-DR3 nor TESS catalogs report its distance.
Flare energies, E$_{\rm bol}$, were calculated following Eq. \ref{ec3}. 
Duration was computed as the difference between the end and start times of the flare event as measured by \textsc{AltaiPony}. Table \ref{limit_flarepar} shows the measured ranges per spectral type for each parameter.

\begin{table}
	\centering
      \caption[]{Measured ranges for flare parameters  per spectral type found in this work.}
        \label{limit_flarepar}     
        \begin{tabular}{lcccc} 
            \hline
            \noalign{\smallskip}
            \multirow{2}{3em}{SpT}   & N$^{\circ}$ &  E$_{\rm bol}$ & Amplitude & Duration \\
                  & (events) & ($\mathrm{erg\,}$)  & (relative flux) & (min) \\
            \noalign{\smallskip}
            \hline
            \noalign{\smallskip}
            M4 & 31  & 2.1$\times10^{30}$ -- 5.58$\times10^{32}$ & 0.004 -- 0.220 & 1 -- 106 \\
            M5 & 354 & 9.9$\times10^{30}$ -- 6.62$\times10^{33}$ & 0.007 -- 3.332 & 1 -- 270 \\
            M6 & 310 & 3.63$\times10^{30}$ -- 1.15$\times10^{34}$ & 0.005 -- 12.457 & 1 -- 240 \\
            M7 & 23  & 1.67$\times10^{31}$ -- 2.49$\times10^{33}$ & 0.04 -- 2.657 & 6 -- 62 \\
            M8 & 36  & 5.26$\times10^{31}$ -- 6.44$\times10^{33}$ & 0.131 -- 15.499 & 6 -- 220 \\
            M9 & 21  & 4.03$\times10^{30}$ -- 2.59$\times10^{33}$ & 0.093 -- 14.016 & 1 -- 60 \\
            L0 & 2  & 1.74$\times10^{32}$ -- 3.30$\times10^{32}$ & 3.764 -- 5.770 & 8 -- 20 \\
            \noalign{\smallskip}
		\hline
	\end{tabular}
\end{table}

\begin{figure}
    \centering
     \includegraphics[width=9.5cm]{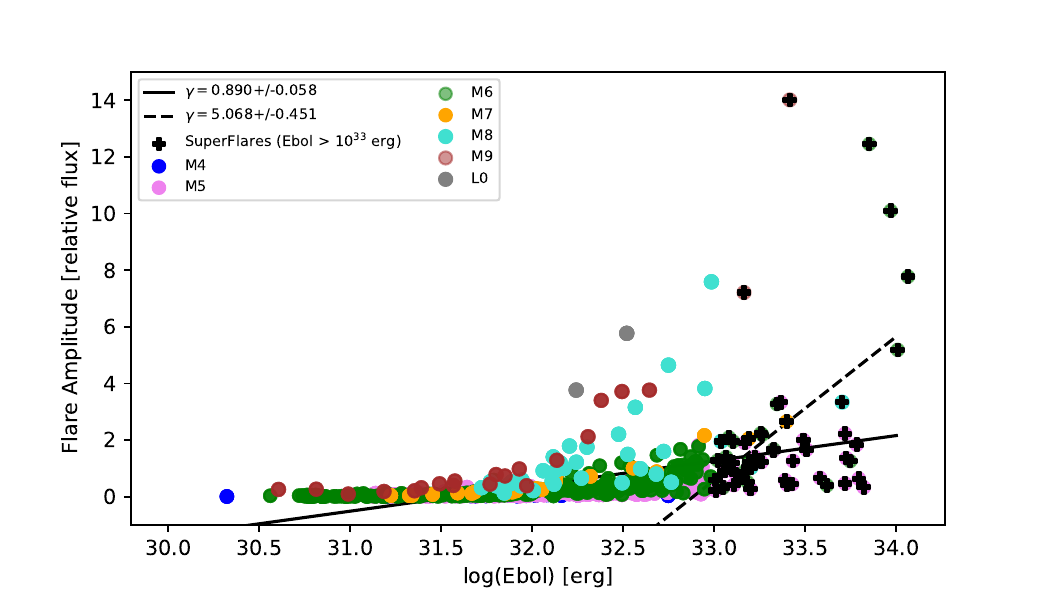}
     \includegraphics[width=9.5cm]{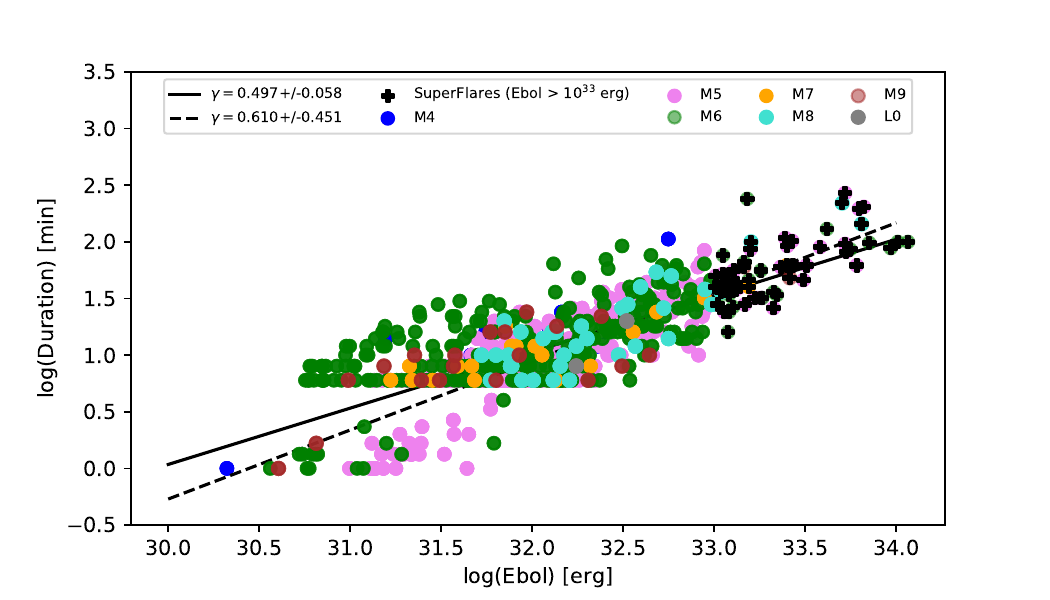}
     \includegraphics[width=9.5cm]{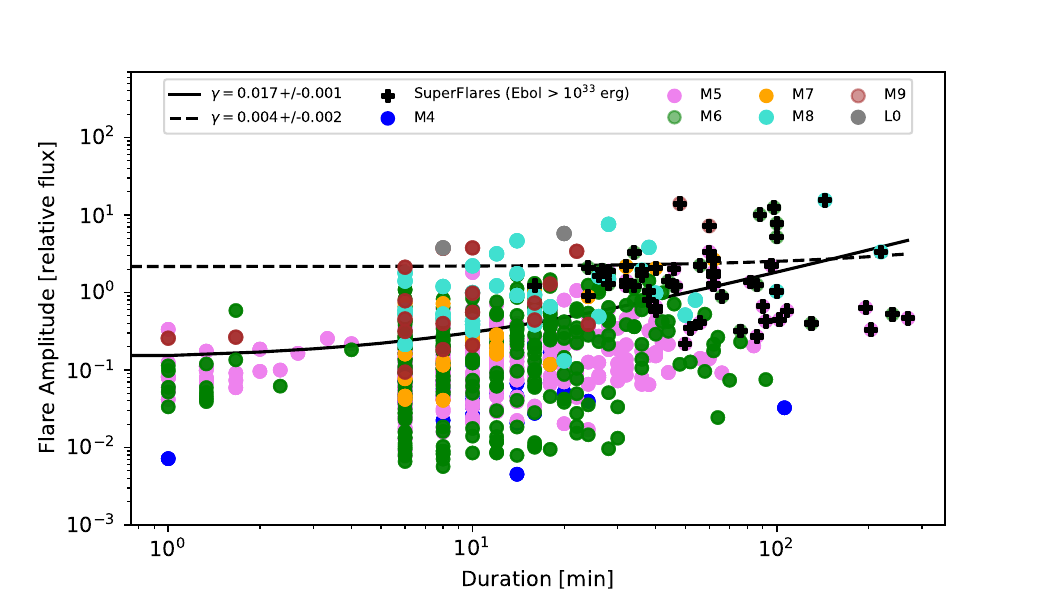}
   \caption{Flare bolometric energy versus flare amplitude (top) and flare duration (middle). Flare duration versus flare amplitude (bottom). Here, blue, pink, green, orange, cyan, brown, and gray circles indicate flares coming from UCDs with M4, M5, M6, M7, M8, M9, and L0 spectral types, respectively, while black crosses point out superflares.
   Solid lines show the best linear least--squares fit to the parameters of all the flares detected,  
   whilst dashed lines represent the best linear least--squares fit but only considering superflares (E$_{\rm bol} > 10^{33}$  $\mathrm{erg\,}$). We note that regardless of spectral type, there are strong correlations between flare energy, amplitude, and duration, showing that higher amplitude flares last longer, and more energetic events, peak higher and last longer than the less energetic ones.
   }
              \label{RelationsFit_FlarePar}
\end{figure}

In Fig. \ref{RelationsFit_FlarePar}, we present plots of flare bolometric energy versus flare amplitude in units of relative flux and flare duration in units of minutes, and flare amplitude against duration. 
Blue, pink, green, orange, cyan, brown, and gray circles indicate flares coming from UCDs with M4, M5, M6, M7, M8, M9, and L0 spectral types, respectively, while black crosses point out superflares.
Regardless of spectral type, strong correlations between energy, amplitude and duration can be seen, showing that higher amplitude flares last longer, and more energetic events, peak higher and last longer than the less energetic ones. This is in agreement with the findings of previous studies. For example, as to the duration-amplitude relation, \cite{hawley2014} and \cite{silverberg2016} used \textit{Kepler} data to measure durations and amplitudes of the flare events in GJ 1243, whilst \cite{raetz2020} did the same for 1644 flares corresponding to 46 K7--M6 stars with detectable rotation period from the K2 short-cadence data. All of them confirmed that flares with higher amplitudes also present longer durations, as found in our study. In contrast, in the recent work of \cite{murray}, the authors do not observe an amplitude-duration relationship for the flares of their M4--L0 targets, with observations taken from ground-based facilities. On the other hand, \cite{raetz2020} noted that the maximum relative flare amplitude increases for later spectral types. As shown in Table \ref{limit_flarepar}, we did not identify the trend from \cite{raetz2020}, probably due to our sample focusing on UCDs and not earlier M-dwarf stars. 

\subsubsection{The energy-duration relation in UCDs is similar to that in partially-convective stars}\label{enerdur}

Additionally, we quantified the relationships between flare parameters. For that purpose, we divided the flares into two groups: one that includes almost all detected flares, and a second group including only superflares (E$_{\rm bol} > 10^{33}\,\mathrm{erg\,}$). 
In Fig. \ref{RelationsFit_FlarePar}, we plot these two groups and the linear least--squares fits to the data. Solid and dashed lines show the best linear fits to the parameters of all the flares detected and to those of the superflares only, respectively. 
 In particular, regarding the energy-duration relationship, \cite{maehara2015} found a slope of $\gamma = 0.39 \pm 0.03$ for solar superflares, which can be explained by assuming magnetic reconnection as the responsible for these events. \cite{silverberg2016} analysed \textit{Kepler} short-cadence data of GJ 1243 and found $\gamma = 0.342 \pm 0.003$  and $\gamma = 0.363 \pm 0.006$ for classical and complex flares, respectively.  Additionally, \cite{yang2023} explored this same correlation for stars of different spectral types and evolutionary states through the analysis of TESS data from the first 30 sectors. Particularly, for M-type stars, they obtained a slope of  0.304 $\pm$ 0.003. 
In this study, we found $\gamma = 0.497 \pm 0.058$ if all the flares are considered and $\gamma = 0.610 \pm 0.451$ for all events with E$_{\rm bol} > 10^{33}\,\mathrm{erg\,}$. 
The agreement in the values of these slopes with those of previous findings might indicate that, although the physical process operating in fully-convective objects in principle differs from that in partially-convective stars, it generates flares of similar characteristics and behaviour than those produced by magnetic reconnection, as in solar-type and early-M stars.

\subsection{Flare Frequency Distribution (FFD)}\label{ffds}

FFD indicates the rate at which a star produces flares above a certain energy. It is represented as a diagram of cumulative flare frequency as a function of flare energy and, it is typically modeled using the following power--law \citep{gershberg1972, lacy1976}:

\begin{equation}
      \log (\nu) = (1-\alpha) \times \log (E_{\rm bol}) + \log(\beta/(1-\alpha)),
\end{equation}

\noindent where $\nu$ represents the number of flares per time unit with energies above a minimum energy, $\mathrm{E_{min}}$, and $\alpha$ and $\beta$ are constants. The value of $\alpha$ is of particular interest because it gives information about the main contributor to the total energy emitted by flares, and hence about the kinds of flares responsible for the coronal heating, during a specific observing window \citep{hudson1991, gudel2003, gao2022}. Specifically, if $\alpha > 2$, low-energy flares supply the majority of the total energy, whilst $\alpha$ $<$ 2, indicates that the high-energy flares have the largest contribution. 

Given the scarce number of flares for M4, M7, and M8--L0 spectral types (31, 23, 36, 21 and 2, respectively) identified in this study, we categorized them in the following groups: one for M4--M5 targets, another for M6--M7,  and the third one for M8--L0.
We constructed the FFDs of all the groups by computing the flare frequency as the ratio of the total number of flares detected to the duration of the TESS sectors in which the objects were observed. 
Previous studies \citep[e.g.][]{gershberg2005, silverberg2016, paudel2018} have shown that, in some cases, FFDs are best fitted with a broken power--law, or a combination of functions, instead of a single power--law. 
In this work, however, also following former studies \citep[e.g.][]{silverberg2016}, in the M4–-M5 group, we did not consider the contribution from flares with $\mathrm{E_{bol}} > 4.6 \times 10^{33}\,\mathrm{erg\,}$ that deviate the FFD from a single power–law.
Meanwhile, in the M8–L0 group, we did not take into account the contribution of the star TIC 318801864 because we were unable to measure the bolometric energy of its only flare (see Section \ref{flarepar} for more details).

\begin{figure}
    \centering
    \includegraphics[width=9.5cm]{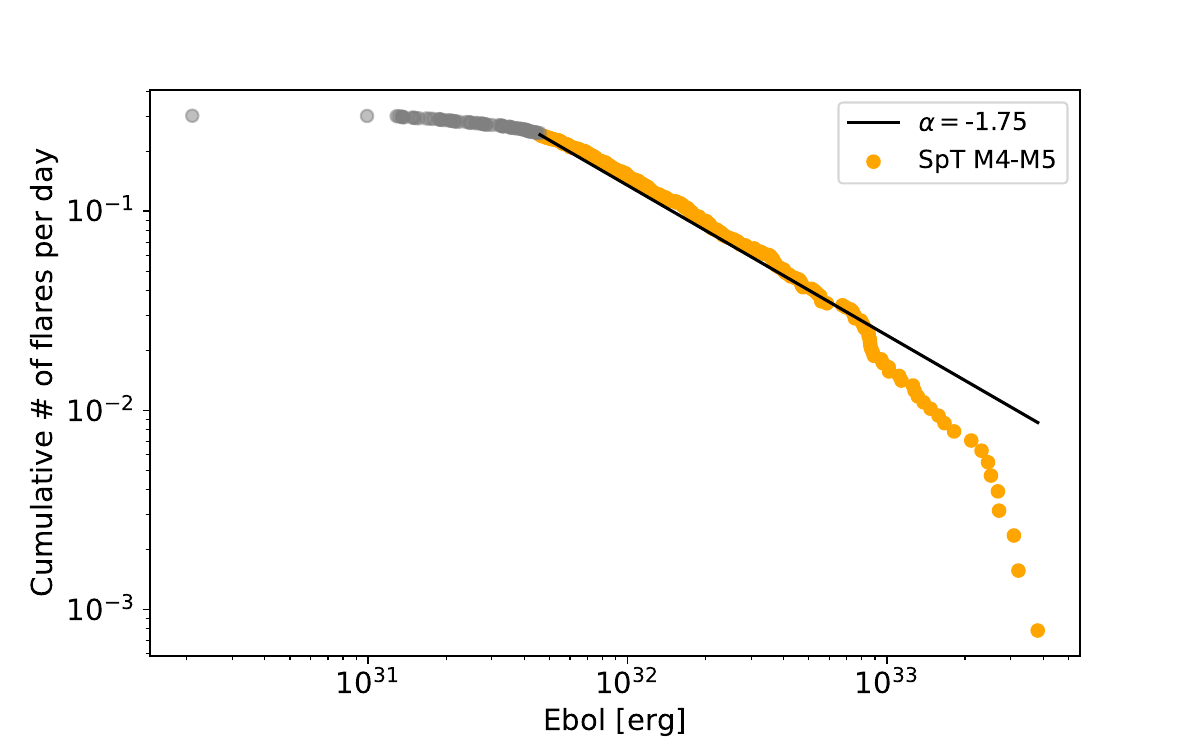}
    \includegraphics[width=9.5cm]{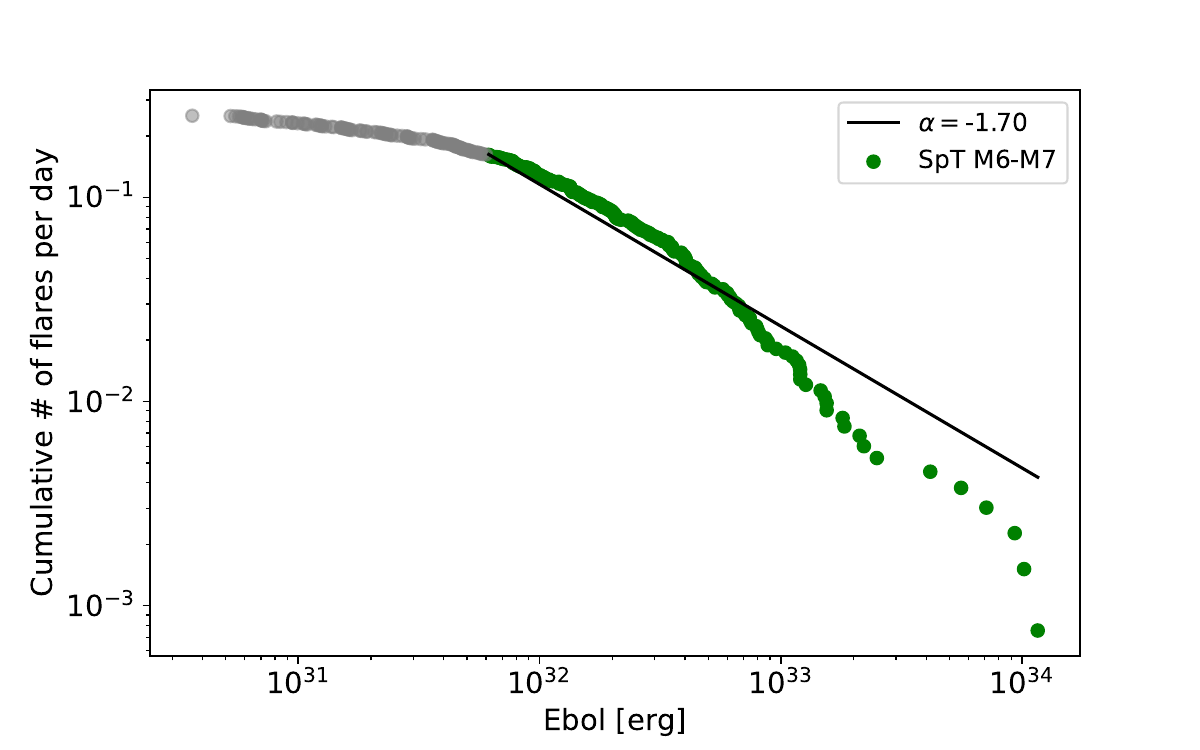}
    \includegraphics[width=9.5cm]{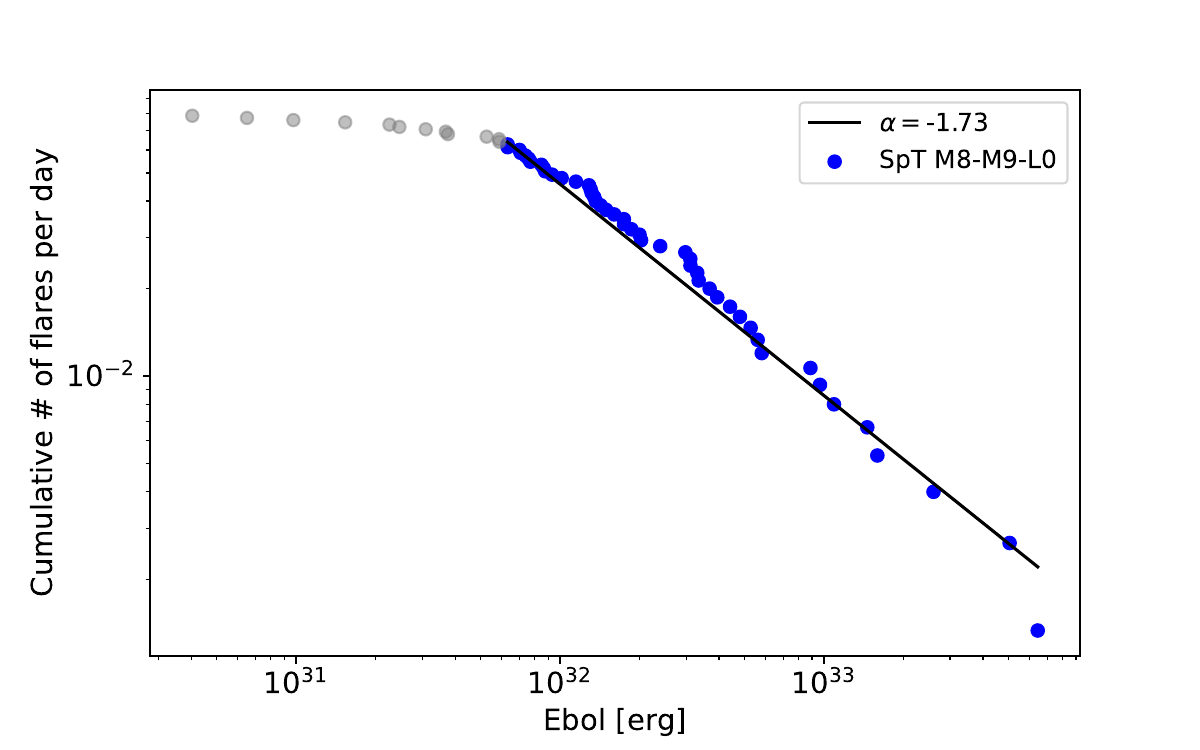}
   \caption{Log-log representation of the FFDs for M4--M5 objects (top in orange), M6--M7 targets (middle in green) and M8--L0 UCDs (bottom in blue).
   In the M4--M5 group, we did not consider the contribution from flares with $\mathrm{E_{bol}} > 4.6 \times 10^{33}\,\mathrm{erg\,}$ that deviate the FFD from a single power–law. Meanwhile, in the M8--L0 group, we did not take into account the flare from the star TIC 318801864 because we were unable to estimate its bolometric energy.
   Black solid lines are the best-fit to the data. Gray symbols indicate flares below the minimum energy value adopted for each spectral type, where the flare distribution is not expected to be complete.
   We found $\alpha = \mathrm{-1.754^{+0.043}_{-0.042}}$, $\alpha = \mathrm{-1.695^{+0.048}_{-0.046}}$ and $\alpha = \mathrm{-1.726^{+0.109}_{-0.100}}$ for the M4--M5, M6--M7, and M8--L0 UCDs, respectively.
   These values are within the range of previous results and indicate that in the UCD regime, there are no changes in the power--law relationship as a function of spectral type.}
              \label{ffd}
\end{figure}

At the low-energy tail, the FFDs show a break in the power--law relationship due to the completeness limit of the sample, i.e. the minimum energy below which the search algorithm is not able to detect all flares, underestimating the frequency. 
Most previous studies applied one of two approaches to handle this issue. One possibility is to compute the minimum energy or the flare recovery rate through artificial flare injection-recovery tests \citep[see][]{seli2021, medina2022, murray} by employing, for example, the tools provided by \textsc{Altaipony}. Alternatively, as implemented in this work, we determined this limiting energy as the minimum energy value for which the slope of the power–law did not vary within the error of the least–squares fit to the data \citep[e.g.][]{hawley2014, silverberg2016}.
In the process, Poisson uncertainties were assigned to the cumulative frequencies to avoid high-energy flares skewing the fit. For M4--M5, M6--M7, and M8--L0 targets,
we found $\mathrm{E_{min}} = 4.6 \times 10^{31}$, $\mathrm{E_{min}} = 6.0 \times 10^{31}$ and $6.0 \times 10^{31}\,\mathrm{erg\,}$, respectively. We applied a Bayesian approach \citep{wheatland2004} provided by \textsc{AltaiPony} to the flares with energies above $\mathrm{E_{min}}$, to determine $\alpha$ and $\beta$ through a Markov Chain Monte Carlo method.
We checked the robustness of the $\alpha$ value determined for the M8--L0 UCDs, given that it was calculated from a small number of flares (only 47). To do so, we recorded the values of the slopes resulting from fitting the FFD several times but removing one flare each time. We found that $\alpha$ remained constant within errors and, hence, the value of the determined slope is robust.
For M4--M5 stars, we obtained $\alpha = \mathrm{-1.753^{+0.043}_{-0.042}}$, $\alpha = \mathrm{-1.695^{+0.048}_{-0.046}}$ for M6--M7 stars, and $\alpha = \mathrm{-1.726^{+0.109}_{-0.100}}$ for M8--L0 UCDs. In all the cases, we used the Kolmogorov--Smirnov statistic \citep{maschberger} to test if the assumption of the power--law hypothesis is correct. We found that the three best-fits are consistent, with a 95$\%$ significance, to a power--law relationship. The resulting FFDs are shown in Fig. \ref{ffd}. 

Several previous works estimated $\alpha$ for UCD FFDs, providing a wide range of possible values. For example, \cite{paudel2018} found $\alpha$ values in the range of 1.3--2.0 for 10 UCDs observed in short-cadence with K2. Additionally, \cite{silverberg2016} obtained $\alpha \sim 2$ for the star GJ 1243 through 11 months of \textit{Kepler} data, but noted a monthly variation of this coefficient from 1.592 to 2.389. \cite{murray} detected flares from 78 low-mass stars observed with the SPECULOOS-Southern Observatory and determined $\alpha$ values of $1.88 \pm 0.05$, $1.72 \pm 0.02$, and $1.82 \pm 0.02$ for M4--M5, M6, and M7 spectral types, respectively. Also, \cite{seli2021} analysed TESS full-frame images of TRAPPIST–1 like ultra-cool dwarfs and found $\alpha = 2.11$. 
In comparison, our values of $\alpha$ for the three groups of UCDs place at the low tail of the distribution. A possible explanation is that the $\mathrm{E_{min}}$ adopted in this work is smaller than the actual minimum limiting energy, producing a less pronounced slope. Nonetheless, more observations of flares with energies below $\mathrm{E_{min}}$ are needed to support this possibility. However, our values are inside the range of previous findings. This agreement within errors in the $\alpha$ values found in this work for the M4--M5, M6--M7 and M8--L0 UCDs, confirms the findings by \cite{murray}, who demonstrated that there are no changes in the power--law relationship as a function of spectral type in the UCD regime.

\subsection{Habitability Potential of M dwarfs hosts}
The potential for habitability of planets around M dwarfs is actively discussed within the astrobiology community \citep[e.g.][]{shieldsetal2016}. The chromospheric activity of these stars may be harmful for habitability, X-ray and extreme UV blow off the planetary atmospheres necessary to retain liquid water at the planet’s surface \citep{doamaral2022}. UV radiation (100--350\,nm) is deemed as harmful to life because it destroys DNA causing mutations and ultimately death, but at the emergence of life UV light was one of the energy sources available for initiating prebiotic chemistry \citep{segura2018}. Recent work has evaluated such potential calculating the amount of UV energy required to drive prebiotic chemistry \citep{rimmer2018} and a sterilization zone where ozone depletion may result in a hostile environment for life at the planetary surface \citep{gunther2020}.

The ozone produced by O$_2$ photolysis protects living organisms from UV damage \citep{segura2018}. Ozone depletion predicted by \citet{segura_effect_2010} and \citet{tilley_modeling_2019} as a result of the combined effect of particles and UV during flares, would happen if the planet atmosphere already had life producing O$_2$. Another possible source of an O$_2$ dominated atmosphere is the catastrophic loss of water predicted for planets around M dwarfs, where the abundance of O$_2$ may exceed 100 bars \citep[e.g.][]{luger_extreme_2015}. The atmospheric chemistry for such atmospheres has not been studied yet, but using the trends calculated in \citet{kozakis_is_2022} is likely to have more O$_3$ with more O$_2$ but its response to the UV from a flare has not been studied yet. Furthermore, the depletion of O$_3$ during flares is mostly caused by the production of NO$x$ by particles, which depends on the abundance of atmospheric N$_2$, which is uncertain. Thus, we cannot make any prediction about the behavior of ozone for these extreme cases of oxygen abundance. In any case, as recognized by \citet{gunther2020}, the lack of an ozone layer is not preventive for the presence of life, thus we do not consider such limits for this discussion. 

\begin{figure}
    \centering
     \includegraphics[width=9.5cm]{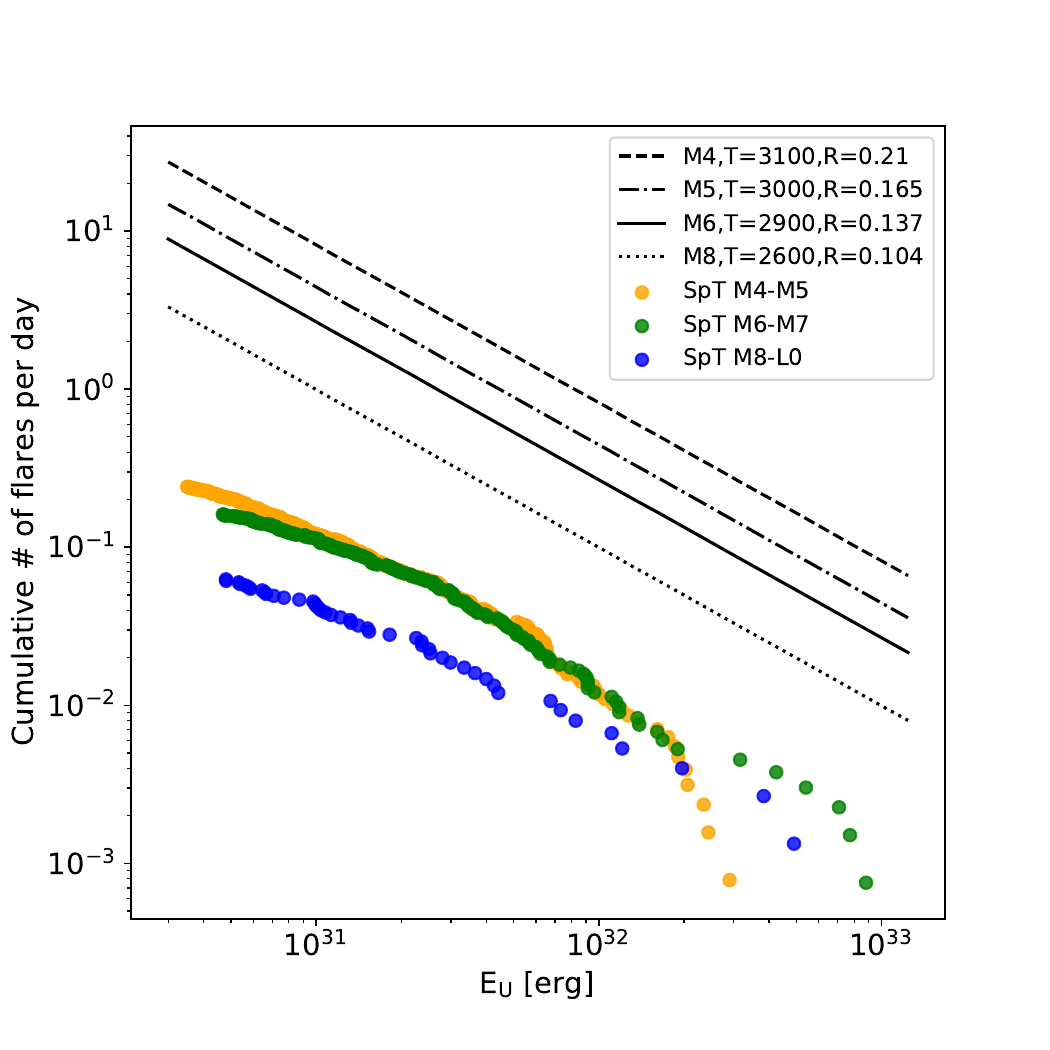}
   \caption{ 
   Abiogenesis zones for the M4--M5, M6--M7 and M8--L0 UCDs studied in this work. The zones were calculated using the stellar parameters of RR Cae (M4, T$_{\rm eff}$=3100 K, R=0.210 R$_{\odot}$), SDSS J0138--0016 (M5, T$_{\rm eff}$=3000 K, R=0.165 R$_{\odot}$), CSS 09704 (M6, T$_{\rm eff}$=2900 K, R=0.137 R$_{\odot}$), and SDSS J0857+0342 (M8, T$_{\rm eff}$=2600 K, R=0.104 R$_{\odot}$) from \citet{parsons2018}. This plot points out that UCDs do not emit enough UV from flares to drive the chemistry of RNA precursors in relevant quantities for the origins of life.}
              \label{habitability}
\end{figure}

The potential for UV to drive chemistry relevant for building RNA precursors was quantified in the `abiogenesis zone' for planets around M dwarfs using their quiescent flux \citet{rimmer2018}. Later, this zone was adapted to consider the UV emitted by flares \citep{gunther2020, ducrot_trappist-1_2020,glazier_evryscope_2020} with the conclusion that neither the quiescent nor the flare UV flux could deliver enough energy to drive prebiotic chemistry, except for a few stars. In Fig. \ref{habitability}, we show the abiogenesis limits from \citet{gunther2020} applied to the results from the previous section. Here, orange, green, and blue circles indicate the FFDs of the M4--M5, M6--M7, and M8--L0 UCDs determined in section \ref{ffds}, but based on the UV energy, E$_{\rm U}$, calculated as  7.6$\%$ of the flare bolometric energy \citep{gunther2020}. Dashed, dashed-dot, solid, and dotted black lines mark the abiogenesis zones that were calculated using the stellar parameters of RR Cae (M4, T$_{\rm eff}$=3100 K, R=0.210 R$_{\odot}$), SDSS J0138--0016 (M5, T$_{\rm eff}$=3000 K, R=0.165 R$_{\odot}$), CSS 09704 (M6, T$_{\rm eff}$=2900 K, R=0.137 R$_{\odot}$), and SDSS J0857+0342 (M8, T$_{\rm eff}$=2600 K, R=0.104 R$_{\odot}$) from \citet{parsons2018}. As can be seen from this plot, and in agreement with previous works \citep[e.g.][]{seli2021, murray}, UCDs do not emit enough UV from flares to drive the chemistry of RNA precursors in relevant quantities for the origins of life.
Although, numerical calculations by \citet{armasvazquez2023} for a known pathway from HCN to adenine –a nucleobase for DNA and RNA-- indicate that large flares produce fast photolysis reactions and the bottleneck to produce compounds relevant for prebiotic chemistry are the kinetic reactions. 

The low UV fluxes from M dwarfs does not prevent their planets from having life, high-energy particles can drive prebiotic chemistry. For early Earth, experiments indicate that galactic cosmic rays 
and solar energetic particles 
may have been the most relevant energy source for molecules relevant for prebiotic chemistry \citep{kobayashi2020}. Proton fluxes accelerated by M dwarfs' flares are expected to be more frequent and intense than those for the Sun \citep{herbst_solar_2019,rodgers-lee_energetic_2023}, therefore they are a potential driver of prebiotic chemistry in potentially habitable planets around these stars. 

\section{Conclusions}\label{conclusions}

In this study, we explored the photometric variability of 208 UCDs, through the analysis of 20-second and 2-minute cadence TESS data. Our main results can be summarised as follows:

\begin{itemize}
    \item We measured rotation periods for 87 objects ($\sim 42\%$) and detected 778 flare events in 103 targets ($\sim 49.5\%$).
    \item No transiting planet or eclipsing binary companion candidate was identified around the targets analysed. 
    \item Around $64\%$ of the UCDs in the sample (i.e. 134 objects) present some indication of activity, either because of the detection of rotational modulation and/or flares.
    \item In terms of rotation and flaring activity, earlier spectral-type UCDs (M4--M6) tend to be more active than later type objects (M7--L4).
    \item No trend was found between rotational period and amplitude and stellar spectral type or effective temperature.
    \item Active UCDs can be found in any of the Milky Way populations (thin disk, thick disk and halo). Noting that the only halo UCD in our sample does not show activity signatures. 
    \item A total of 56 superflares with bolometric energies between $1.0 \times 10^{33}$ and $1.1 \times 10^{34} \mathrm{erg\,}$ from 33 UCDs were detected.
    \item For all spectral types, strong correlations between bolometric energy, amplitude, and duration of flares can be seen, showing that higher amplitude flares last longer, and more energetic events peak higher and last longer than the less energetic ones.
    \item For the flare energy-duration correlation, we found a slope of $\gamma = 0.497 \pm 0.058$ if all the flares are considered and $\gamma = 0.610 \pm 0.451$ for superflare events, both are in agreement, given the uncertainties, with the results of previous studies for solar-type and earlier M dwarfs.
   \item 
   The slope of the FFD for M4--M5 UCDs is measured to be  $\alpha = \mathrm{-1.75 \pm 0.04}$, for  M6--M7 UCDs is  $\alpha = \mathrm{-1.69 \pm 0.04}$, and for M8--L0 UCDs is $\alpha = \mathrm{-1.72 \pm 0.1}$, and confirms
   previous findings demonstrating that there are no changes in the power--law relationship as a function of spectral type in the UCD regime.
   \item UV radiation from the flares of the UCDs analysed in this work may not be enough to drive prebiotic chemistry. However, high-energy particles have the potential to start such chemistry considering the higher CO abundances that terrestrial atmosphere could develope around M dwarfs. 
\end{itemize}

Although the dynamo mechanism dominating the interiors of UCDs must differ from the $\alpha\omega$ dynamo operating in stars with tachoclines, most of these findings show that the signatures of magnetic activity, such as flares and rotational modulation, have similar characteristics among partially-convective FGK and M stars and fully-convective UCDs.

\section*{Acknowledgements}
The authors thank the anonymous referee for a thoughtful reading of the manuscript and for providing very constructive comments and corrections that certainly improved the scientific quality of this paper. This work has been partially supported by UNAM-PAPIIT-IG101321 and PIBAA-CONICET ID-73811. We acknowledge the use of public TESS data from pipelines at the TESS Science Office and at the TESS Science Processing Operations Center. R.P. and E.J. thank L. Messi and L. Scaloni for inspiration and motivation through their continuous examples of humbleness, perseverance, sacrifice, and hard work. Data presented in this paper were obtained from the Mikulski Archive for Space Telescopes (MAST). This work has made use of data from the European Space Agency (ESA) mission \Gaia\ (\url{https://www.cosmos.esa.int/gaia}), processed by the \Gaia\ Data Processing and Analysis Consortium (DPAC, \url{https://www.cosmos.esa.int/web/gaia/dpac/consortium}). Funding for the DPAC has been provided by national institutions, in particular the institutions participating in the \Gaia\ Multilateral Agreement.
This research has made use of the SIMBAD database, operated at CDS, Strasbourg, France. This research has made use of "Aladin sky atlas" developed at CDS, Strasbourg Observatory, France.
This work made use of \textsc{tpfplotter} by J. Lillo-Box (publicly available in www.github.com/jlillo/tpfplotter), which also made use of the python packages \texttt{astropy}, \texttt{lightkurve},
\texttt{matplotlib} and \texttt{numpy}.

\section*{Data Availability}

The TESS data is accessible via the MAST (Mikulski Archive for Space Telescopes) portal at https://mast.stsci.edu/portal/Mashup/Clients/Mast/Portal.html.




\bibliographystyle{mnras}
\bibliography{Petrucci-2023-UCDs-MNRAS} 

\begin{thebibliography}{}
\makeatletter
\relax
\def\mn@urlcharsother{\let\do\@makeother \do\$\do\&\do\#\do\^\do\_\do\%\do\~}
\def\mn@doi{\begingroup\mn@urlcharsother \@ifnextchar [ {\mn@doi@} {\mn@doi@[]}}
\def\mn@doi@[#1]#2{\def\@tempa{#1}\ifx\@tempa\@empty \href {http://dx.doi.org/#2} {doi:#2}\else \href {http://dx.doi.org/#2} {#1}\fi \endgroup}
\def\mn@eprint#1#2{\mn@eprint@#1:#2::\@nil}
\def\mn@eprint@arXiv#1{\href {http://arxiv.org/abs/#1} {{\tt arXiv:#1}}}
\def\mn@eprint@dblp#1{\href {http://dblp.uni-trier.de/rec/bibtex/#1.xml} {dblp:#1}}
\def\mn@eprint@#1:#2:#3:#4\@nil{\def\@tempa {#1}\def\@tempb {#2}\def\@tempc {#3}\ifx \@tempc \@empty \let \@tempc \@tempb \let \@tempb \@tempa \fi \ifx \@tempb \@empty \def\@tempb {arXiv}\fi \@ifundefined {mn@eprint@\@tempb}{\@tempb:\@tempc}{\expandafter \expandafter \csname mn@eprint@\@tempb\endcsname \expandafter{\@tempc}}}

\bibitem[\protect\citeauthoryear{{Affer}, {Micela}, {Favata}  \& {Flaccomio}}{{Affer} et~al.}{2012}]{affer2012}
{Affer} L.,  {Micela} G.,  {Favata} F.,   {Flaccomio} E.,  2012, \mn@doi [\mnras] {10.1111/j.1365-2966.2012.20802.x}, \href {https://ui.adsabs.harvard.edu/abs/2012MNRAS.424...11A} {424, 11}

\bibitem[\protect\citeauthoryear{{Alibert} \& {Benz}}{{Alibert} \& {Benz}}{2017}]{alibert2017}
{Alibert} Y.,  {Benz} W.,  2017, \mn@doi [\aap] {10.1051/0004-6361/201629671}, \href {https://ui.adsabs.harvard.edu/abs/2017A&A...598L...5A} {598, L5}

\bibitem[\protect\citeauthoryear{{Andrews}, {Rosenfeld}, {Kraus}  \& {Wilner}}{{Andrews} et~al.}{2013}]{andrews2013}
{Andrews} S.~M.,  {Rosenfeld} K.~A.,  {Kraus} A.~L.,   {Wilner} D.~J.,  2013, \mn@doi [\apj] {10.1088/0004-637X/771/2/129}, \href {https://ui.adsabs.harvard.edu/abs/2013ApJ...771..129A} {771, 129}

\bibitem[\protect\citeauthoryear{{Anglada-Escud{\'e}} et~al.,}{{Anglada-Escud{\'e}} et~al.}{2016}]{anglada-escude2016}
{Anglada-Escud{\'e}} G.,  et~al., 2016, \mn@doi [\nat] {10.1038/nature19106}, \href {https://ui.adsabs.harvard.edu/abs/2016Natur.536..437A} {536, 437}

\bibitem[\protect\citeauthoryear{{Angus} et~al.,}{{Angus} et~al.}{2020}]{angus2020}
{Angus} R.,  et~al., 2020, \mn@doi [\aj] {10.3847/1538-3881/ab91b2}, \href {https://ui.adsabs.harvard.edu/abs/2020AJ....160...90A} {160, 90}

\bibitem[\protect\citeauthoryear{{Anthony} et~al.,}{{Anthony} et~al.}{2022}]{anthony2022}
{Anthony} F.,  et~al., 2022, \mn@doi [\aj] {10.3847/1538-3881/ac6110}, \href {https://ui.adsabs.harvard.edu/abs/2022AJ....163..257A} {163, 257}

\bibitem[\protect\citeauthoryear{{Armas-V{\'a}zquez}, {Gonz{\'a}lez-Espinoza}, {Segura}, {Heredia}  \& {Miranda-Rosete}}{{Armas-V{\'a}zquez} et~al.}{2023}]{armasvazquez2023}
{Armas-V{\'a}zquez} M.~Z.,  {Gonz{\'a}lez-Espinoza} C.~E.,  {Segura} A.,  {Heredia} A.,   {Miranda-Rosete} A.,  2023, \mn@doi [Astrobiology] {10.1089/ast.2022.0050}, \href {https://ui.adsabs.harvard.edu/abs/2023AsBio..23..705A} {23, 705}

\bibitem[\protect\citeauthoryear{{Baluev}}{{Baluev}}{2008}]{baluev2008}
{Baluev} R.~V.,  2008, \mn@doi [\mnras] {10.1111/j.1365-2966.2008.12689.x}, \href {https://ui.adsabs.harvard.edu/abs/2008MNRAS.385.1279B} {385, 1279}

\bibitem[\protect\citeauthoryear{{Bolmont}, {Selsis}, {Owen}, {Ribas}, {Raymond}, {Leconte}  \& {Gillon}}{{Bolmont} et~al.}{2017}]{bolmont2017}
{Bolmont} E.,  {Selsis} F.,  {Owen} J.~E.,  {Ribas} I.,  {Raymond} S.~N.,  {Leconte} J.,   {Gillon} M.,  2017, \mn@doi [\mnras] {10.1093/mnras/stw2578}, \href {https://ui.adsabs.harvard.edu/abs/2017MNRAS.464.3728B} {464, 3728}

\bibitem[\protect\citeauthoryear{{Bonfils} et~al.,}{{Bonfils} et~al.}{2015}]{shaklan2015}
{Bonfils} X.,  et~al., 2015, in {Shaklan} S.,  ed.,  Society of Photo-Optical Instrumentation Engineers (SPIE) Conference Series Vol. 9605, Techniques and Instrumentation for Detection of Exoplanets VII. p. 96051L (\mn@eprint {arXiv} {1508.06601}), \mn@doi{10.1117/12.2186999}

\bibitem[\protect\citeauthoryear{{Borucki} et~al.,}{{Borucki} et~al.}{2010}]{borucki2010}
{Borucki} W.~J.,  et~al., 2010, \mn@doi [Science] {10.1126/science.1185402}, \href {https://ui.adsabs.harvard.edu/abs/2010Sci...327..977B} {327, 977}

\bibitem[\protect\citeauthoryear{{Bouvier}}{{Bouvier}}{2007}]{bouvier2007}
{Bouvier} J.,  2007, in {Bouvier} J.,  {Appenzeller} I.,  eds, ~ Vol. 243, Star-Disk Interaction in Young Stars. pp 231--240 (\mn@eprint {arXiv} {0712.2988}), \mn@doi{10.1017/S1743921307009593}

\bibitem[\protect\citeauthoryear{{Browning}}{{Browning}}{2008}]{browning2008}
{Browning} M.~K.,  2008, \mn@doi [\apj] {10.1086/527432}, \href {https://ui.adsabs.harvard.edu/abs/2008ApJ...676.1262B} {676, 1262}

\bibitem[\protect\citeauthoryear{{Chabrier} \& {K{\"u}ker}}{{Chabrier} \& {K{\"u}ker}}{2006}]{chabrier2006}
{Chabrier} G.,  {K{\"u}ker} M.,  2006, \mn@doi [\aap] {10.1051/0004-6361:20042475}, \href {https://ui.adsabs.harvard.edu/abs/2006A&A...446.1027C} {446, 1027}

\bibitem[\protect\citeauthoryear{{Chang}, {Byun}  \& {Hartman}}{{Chang} et~al.}{2015}]{chang2015}
{Chang} S.~W.,  {Byun} Y.~I.,   {Hartman} J.~D.,  2015, \mn@doi [\apj] {10.1088/0004-637X/814/1/35}, \href {https://ui.adsabs.harvard.edu/abs/2015ApJ...814...35C} {814, 35}

\bibitem[\protect\citeauthoryear{{Charbonneau}}{{Charbonneau}}{2010}]{charbonneau2010}
{Charbonneau} P.,  2010, \mn@doi [Living Reviews in Solar Physics] {10.12942/lrsp-2010-3}, \href {https://ui.adsabs.harvard.edu/abs/2010LRSP....7....3C} {7, 3}

\bibitem[\protect\citeauthoryear{{Climent}, {Guirado}, {P{\'e}rez-Torres}, {Marcaide}  \& {Pe{\~n}a-Mo{\~n}ino}}{{Climent} et~al.}{2023}]{climent2023}
{Climent} J.~B.,  {Guirado} J.~C.,  {P{\'e}rez-Torres} M.,  {Marcaide} J.~M.,   {Pe{\~n}a-Mo{\~n}ino} L.,  2023, \mn@doi [Science] {10.1126/science.adg6635}, \href {https://ui.adsabs.harvard.edu/abs/2023Sci...381.1120C} {381, 1120}

\bibitem[\protect\citeauthoryear{{Cody} \& {Hillenbrand}}{{Cody} \& {Hillenbrand}}{2010}]{cody2010}
{Cody} A.~M.,  {Hillenbrand} L.~A.,  2010, \mn@doi [\apjs] {10.1088/0067-0049/191/2/389}, \href {https://ui.adsabs.harvard.edu/abs/2010ApJS..191..389C} {191, 389}

\bibitem[\protect\citeauthoryear{{Cody}, {Hillenbrand}  \& {Rebull}}{{Cody} et~al.}{2022}]{cody2022}
{Cody} A.~M.,  {Hillenbrand} L.~A.,   {Rebull} L.~M.,  2022, \mn@doi [\aj] {10.3847/1538-3881/ac5b73}, \href {https://ui.adsabs.harvard.edu/abs/2022AJ....163..212C} {163, 212}

\bibitem[\protect\citeauthoryear{{Davenport}}{{Davenport}}{2016}]{davenport2016}
{Davenport} J. R.~A.,  2016, \mn@doi [\apj] {10.3847/0004-637X/829/1/23}, \href {https://ui.adsabs.harvard.edu/abs/2016ApJ...829...23D} {829, 23}

\bibitem[\protect\citeauthoryear{{Delrez} et~al.,}{{Delrez} et~al.}{2018}]{delrez2018}
{Delrez} L.,  et~al., 2018, in {Marshall} H.~K.,  {Spyromilio} J.,  eds,  Society of Photo-Optical Instrumentation Engineers (SPIE) Conference Series Vol. 10700, Ground-based and Airborne Telescopes VII. p. 107001I (\mn@eprint {arXiv} {1806.11205}), \mn@doi{10.1117/12.2312475}

\bibitem[\protect\citeauthoryear{{Donati} et~al.,}{{Donati} et~al.}{2018}]{donati2018}
{Donati} J.-F.,  et~al., 2018, in {Deeg} H.~J.,  {Belmonte} J.~A.,  eds, , Handbook of Exoplanets.
p.~107, \mn@doi{10.1007/978-3-319-55333-7_107}

\bibitem[\protect\citeauthoryear{Ducrot et~al.,}{Ducrot et~al.}{2020}]{ducrot_trappist-1_2020}
Ducrot E.,  et~al., 2020, \mn@doi [Astronomy \& Astrophysics] {10.1051/0004-6361/201937392}, 640, A112

\bibitem[\protect\citeauthoryear{{Flores}, {Connelley}, {Reipurth}  \& {Boogert}}{{Flores} et~al.}{2019}]{flores2019}
{Flores} C.,  {Connelley} M.~S.,  {Reipurth} B.,   {Boogert} A.,  2019, \mn@doi [\apj] {10.3847/1538-4357/ab35d4}, \href {https://ui.adsabs.harvard.edu/abs/2019ApJ...882...75F} {882, 75}

\bibitem[\protect\citeauthoryear{{Gaia Collaboration} et~al.,}{{Gaia Collaboration} et~al.}{2016}]{gaia2016}
{Gaia Collaboration} et~al., 2016, \mn@doi [\aap] {10.1051/0004-6361/201629272}, \href {https://ui.adsabs.harvard.edu/abs/2016A&A...595A...1G} {595, A1}

\bibitem[\protect\citeauthoryear{{Gaia Collaboration} et~al.,}{{Gaia Collaboration} et~al.}{2023}]{gaia2022}
{Gaia Collaboration} et~al., 2023, \mn@doi [\aap] {10.1051/0004-6361/202243940}, \href {https://ui.adsabs.harvard.edu/abs/2023A&A...674A...1G} {674, A1}

\bibitem[\protect\citeauthoryear{{Gao}, {Liu}, {Yang}  \& {Zhou}}{{Gao} et~al.}{2022}]{gao2022}
{Gao} D.-Y.,  {Liu} H.-G.,  {Yang} M.,   {Zhou} J.-L.,  2022, \mn@doi [\aj] {10.3847/1538-3881/ac937e}, \href {https://ui.adsabs.harvard.edu/abs/2022AJ....164..213G} {164, 213}

\bibitem[\protect\citeauthoryear{{Gardner} et~al.,}{{Gardner} et~al.}{2006}]{gardner2006}
{Gardner} J.~P.,  et~al., 2006, in {Mather} J.~C.,  {MacEwen} H.~A.,   {de Graauw} M. W.~M.,  eds,  Society of Photo-Optical Instrumentation Engineers (SPIE) Conference Series Vol. 6265, Society of Photo-Optical Instrumentation Engineers (SPIE) Conference Series. p. 62650N, \mn@doi{10.1117/12.670492}

\bibitem[\protect\citeauthoryear{{Gastine}, {Morin}, {Duarte}, {Reiners}, {Christensen}  \& {Wicht}}{{Gastine} et~al.}{2013}]{gastine2013}
{Gastine} T.,  {Morin} J.,  {Duarte} L.,  {Reiners} A.,  {Christensen} U.~R.,   {Wicht} J.,  2013, \mn@doi [\aap] {10.1051/0004-6361/201220317}, \href {https://ui.adsabs.harvard.edu/abs/2013A&A...549L...5G} {549, L5}

\bibitem[\protect\citeauthoryear{{Gershberg}}{{Gershberg}}{1972}]{gershberg1972}
{Gershberg} R.~E.,  1972, \mn@doi [\apss] {10.1007/BF00643168}, \href {https://ui.adsabs.harvard.edu/abs/1972Ap&SS..19...75G} {19, 75}

\bibitem[\protect\citeauthoryear{{Gershberg}}{{Gershberg}}{2005}]{gershberg2005}
{Gershberg} R.~E.,  2005, {Solar-Type Activity in Main-Sequence Stars}, \mn@doi{10.1007/3-540-28243-2.
}

\bibitem[\protect\citeauthoryear{{Getman}, {Feigelson}, {Garmire}, {Broos}, {Kuhn}, {Preibisch}  \& {Airapetian}}{{Getman} et~al.}{2022}]{getman2022}
{Getman} K.~V.,  {Feigelson} E.~D.,  {Garmire} G.~P.,  {Broos} P.~S.,  {Kuhn} M.~A.,  {Preibisch} T.,   {Airapetian} V.~S.,  2022, \mn@doi [\apj] {10.3847/1538-4357/ac7c69}, \href {https://ui.adsabs.harvard.edu/abs/2022ApJ...935...43G} {935, 43}

\bibitem[\protect\citeauthoryear{{Getman}, {Feigelson}  \& {Garmire}}{{Getman} et~al.}{2023}]{getman2023}
{Getman} K.~V.,  {Feigelson} E.~D.,   {Garmire} G.~P.,  2023, \mn@doi [\apj] {10.3847/1538-4357/acd690}, \href {https://ui.adsabs.harvard.edu/abs/2023ApJ...952...63G} {952, 63}

\bibitem[\protect\citeauthoryear{{Gillon} et~al.,}{{Gillon} et~al.}{2016}]{gillon2016}
{Gillon} M.,  et~al., 2016, \mn@doi [\nat] {10.1038/nature17448}, \href {https://ui.adsabs.harvard.edu/abs/2016Natur.533..221G} {533, 221}

\bibitem[\protect\citeauthoryear{{Gillon} et~al.,}{{Gillon} et~al.}{2017}]{gillon2017}
{Gillon} M.,  et~al., 2017, \mn@doi [\nat] {10.1038/nature21360}, \href {https://ui.adsabs.harvard.edu/abs/2017Natur.542..456G} {542, 456}

\bibitem[\protect\citeauthoryear{Glazier, Howard, Corbett, Law, Ratzloff, Fors  \& Ser}{Glazier et~al.}{2020}]{glazier_evryscope_2020}
Glazier A.~L.,  Howard W.~S.,  Corbett H.,  Law N.~M.,  Ratzloff J.~K.,  Fors O.,   Ser D.~d.,  2020, \mn@doi [The Astrophysical Journal] {10.3847/1538-4357/aba4a6}, 900, 27

\bibitem[\protect\citeauthoryear{{G{\'o}mez Maqueo Chew} et~al.,}{{G{\'o}mez Maqueo Chew} et~al.}{2023}]{GomezMaqueo2023}
{G{\'o}mez Maqueo Chew} Y.,  et~al., 2023, in Revista Mexicana de Astronomia y Astrofisica Conference Series. pp 44--46, \mn@doi{https://doi.org/10.22201/ia.14052059p.2023.55.10}

\bibitem[\protect\citeauthoryear{{G{\"u}del}, {Audard}, {Kashyap}, {Drake}  \& {Guinan}}{{G{\"u}del} et~al.}{2003}]{gudel2003}
{G{\"u}del} M.,  {Audard} M.,  {Kashyap} V.~L.,  {Drake} J.~J.,   {Guinan} E.~F.,  2003, \mn@doi [\apj] {10.1086/344614}, \href {https://ui.adsabs.harvard.edu/abs/2003ApJ...582..423G} {582, 423}

\bibitem[\protect\citeauthoryear{{G{\"u}nther} et~al.,}{{G{\"u}nther} et~al.}{2020}]{gunther2020}
{G{\"u}nther} M.~N.,  et~al., 2020, \mn@doi [\aj] {10.3847/1538-3881/ab5d3a}, \href {https://ui.adsabs.harvard.edu/abs/2020AJ....159...60G} {159, 60}

\bibitem[\protect\citeauthoryear{{Hawley}, {Davenport}, {Kowalski}, {Wisniewski}, {Hebb}, {Deitrick}  \& {Hilton}}{{Hawley} et~al.}{2014}]{hawley2014}
{Hawley} S.~L.,  {Davenport} J. R.~A.,  {Kowalski} A.~F.,  {Wisniewski} J.~P.,  {Hebb} L.,  {Deitrick} R.,   {Hilton} E.~J.,  2014, \mn@doi [\apj] {10.1088/0004-637X/797/2/121}, \href {https://ui.adsabs.harvard.edu/abs/2014ApJ...797..121H} {797, 121}

\bibitem[\protect\citeauthoryear{{Herbst}, {Bailer-Jones}, {Mundt}, {Meisenheimer}  \& {Wackermann}}{{Herbst} et~al.}{2002}]{herbst2002}
{Herbst} W.,  {Bailer-Jones} C.~A.~L.,  {Mundt} R.,  {Meisenheimer} K.,   {Wackermann} R.,  2002, \mn@doi [\aap] {10.1051/0004-6361:20021362}, \href {https://ui.adsabs.harvard.edu/abs/2002A&A...396..513H} {396, 513}

\bibitem[\protect\citeauthoryear{Herbst, Papaioannou, Banjac  \& Heber}{Herbst et~al.}{2019}]{herbst_solar_2019}
Herbst K.,  Papaioannou A.,  Banjac S.,   Heber B.,  2019, \mn@doi [Astronomy \& Astrophysics] {10.1051/0004-6361/201832789}, 621, A67

\bibitem[\protect\citeauthoryear{Hippke \& Heller}{Hippke \& Heller}{2019}]{hippke2019b}
Hippke M.,  Heller R.,  2019, \mn@doi [Astronomy and Astrophysics] {10.1051/0004-6361/201834672}, 623, A39

\bibitem[\protect\citeauthoryear{{Hippke}, {David}, {Mulders}  \& {Heller}}{{Hippke} et~al.}{2019}]{hippke2019a}
{Hippke} M.,  {David} T.~J.,  {Mulders} G.~D.,   {Heller} R.,  2019, \mn@doi [\aj] {10.3847/1538-3881/ab3984}, \href {https://ui.adsabs.harvard.edu/abs/2019AJ....158..143H} {158, 143}

\bibitem[\protect\citeauthoryear{{Howard}}{{Howard}}{2022}]{howard2022}
{Howard} W.~S.,  2022, \mn@doi [\mnras] {10.1093/mnrasl/slac024}, \href {https://ui.adsabs.harvard.edu/abs/2022MNRAS.512L..60H} {512, L60}

\bibitem[\protect\citeauthoryear{{Howard} \& {MacGregor}}{{Howard} \& {MacGregor}}{2022}]{howardmcgregor2022}
{Howard} W.~S.,  {MacGregor} M.~A.,  2022, \mn@doi [\apj] {10.3847/1538-4357/ac426e}, \href {https://ui.adsabs.harvard.edu/abs/2022ApJ...926..204H} {926, 204}

\bibitem[\protect\citeauthoryear{{Hudson}}{{Hudson}}{1991}]{hudson1991}
{Hudson} H.~S.,  1991, \mn@doi [\solphys] {10.1007/BF00149894}, \href {https://ui.adsabs.harvard.edu/abs/1991SoPh..133..357H} {133, 357}

\bibitem[\protect\citeauthoryear{{Ilin}, {Schmidt}, {Poppenh{\"a}ger}, {Davenport}, {Kristiansen}  \& {Omohundro}}{{Ilin} et~al.}{2021}]{ilin2021}
{Ilin} E.,  {Schmidt} S.~J.,  {Poppenh{\"a}ger} K.,  {Davenport} J. R.~A.,  {Kristiansen} M.~H.,   {Omohundro} M.,  2021, \mn@doi [\aap] {10.1051/0004-6361/202039198}, \href {https://ui.adsabs.harvard.edu/abs/2021A&A...645A..42I} {645, A42}

\bibitem[\protect\citeauthoryear{{Irwin}, {Charbonneau}, {Nutzman}  \& {Falco}}{{Irwin} et~al.}{2009}]{irwin2009}
{Irwin} J.,  {Charbonneau} D.,  {Nutzman} P.,   {Falco} E.,  2009, in {Pont} F.,  {Sasselov} D.,   {Holman} M.~J.,  eds, ~ Vol. 253, Transiting Planets. pp 37--43 (\mn@eprint {arXiv} {0807.1316}), \mn@doi{10.1017/S1743921308026215}

\bibitem[\protect\citeauthoryear{{Jackman} et~al.,}{{Jackman} et~al.}{2021}]{jackman2021}
{Jackman} J. A.~G.,  et~al., 2021, \mn@doi [\mnras] {10.1093/mnras/stab979}, \href {https://ui.adsabs.harvard.edu/abs/2021MNRAS.504.3246J} {504, 3246}

\bibitem[\protect\citeauthoryear{{Jenkins} et~al.,}{{Jenkins} et~al.}{2016}]{jenkins2016}
{Jenkins} J.~M.,  et~al., 2016, in {Chiozzi} G.,  {Guzman} J.~C.,  eds,  Society of Photo-Optical Instrumentation Engineers (SPIE) Conference Series Vol. 9913, Software and Cyberinfrastructure for Astronomy IV. p. 99133E, \mn@doi{10.1117/12.2233418}

\bibitem[\protect\citeauthoryear{{Johnson} \& {Soderblom}}{{Johnson} \& {Soderblom}}{1987}]{johnson1987}
{Johnson} D. R.~H.,  {Soderblom} D.~R.,  1987, \mn@doi [\aj] {10.1086/114370}, \href {https://ui.adsabs.harvard.edu/abs/1987AJ.....93..864J} {93, 864}

\bibitem[\protect\citeauthoryear{{Katz} et~al.,}{{Katz} et~al.}{2023}]{katz2022}
{Katz} D.,  et~al., 2023, \mn@doi [\aap] {10.1051/0004-6361/202244220}, \href {https://ui.adsabs.harvard.edu/abs/2023A&A...674A...5K} {674, A5}

\bibitem[\protect\citeauthoryear{{Kirkpatrick}, {Henry}  \& {Simons}}{{Kirkpatrick} et~al.}{1995}]{kirkpatick1995}
{Kirkpatrick} J.~D.,  {Henry} T.~J.,   {Simons} D.~A.,  1995, \mn@doi [\aj] {10.1086/117323}, \href {https://ui.adsabs.harvard.edu/abs/1995AJ....109..797K} {109, 797}

\bibitem[\protect\citeauthoryear{Kobayashi et~al.,}{Kobayashi et~al.}{2023}]{kobayashi2020}
Kobayashi K.,  et~al., 2023, \mn@doi [Life] {10.3390/life13051103}, 13

\bibitem[\protect\citeauthoryear{{Kochukhov}}{{Kochukhov}}{2021}]{kochukhov2021}
{Kochukhov} O.,  2021, \mn@doi [\aapr] {10.1007/s00159-020-00130-3}, \href {https://ui.adsabs.harvard.edu/abs/2021A&ARv..29....1K} {29, 1}

\bibitem[\protect\citeauthoryear{Kozakis, Mendonça  \& Buchhave}{Kozakis et~al.}{2022}]{kozakis_is_2022}
Kozakis T.,  Mendonça J.~M.,   Buchhave L.~A.,  2022, \mn@doi [Astronomy \& Astrophysics] {10.1051/0004-6361/202244164}, 665, A156

\bibitem[\protect\citeauthoryear{{Lacy}, {Moffett}  \& {Evans}}{{Lacy} et~al.}{1976}]{lacy1976}
{Lacy} C.~H.,  {Moffett} T.~J.,   {Evans} D.~S.,  1976, \mn@doi [\apjs] {10.1086/190358}, \href {https://ui.adsabs.harvard.edu/abs/1976ApJS...30...85L} {30, 85}

\bibitem[\protect\citeauthoryear{{Lamm}, {Mundt}, {Bailer-Jones}  \& {Herbst}}{{Lamm} et~al.}{2005}]{lamm2005}
{Lamm} M.~H.,  {Mundt} R.,  {Bailer-Jones} C.~A.~L.,   {Herbst} W.,  2005, \mn@doi [\aap] {10.1051/0004-6361:20040492}, \href {https://ui.adsabs.harvard.edu/abs/2005A&A...430.1005L} {430, 1005}

\bibitem[\protect\citeauthoryear{{Lightkurve Collaboration} et~al.,}{{Lightkurve Collaboration} et~al.}{2018}]{lightkurve2018}
{Lightkurve Collaboration} et~al., 2018, {Lightkurve: Kepler and TESS time series analysis in Python}, Astrophysics Source Code Library (\mn@eprint {ascl} {1812.013})

\bibitem[\protect\citeauthoryear{{Lomb}}{{Lomb}}{1976}]{lomb1976}
{Lomb} N.~R.,  1976, \mn@doi [\apss] {10.1007/BF00648343}, \href {https://ui.adsabs.harvard.edu/abs/1976Ap&SS..39..447L} {39, 447}

\bibitem[\protect\citeauthoryear{{L{\'o}pez-Valdivia} et~al.,}{{L{\'o}pez-Valdivia} et~al.}{2023}]{lopezvaldivia2023}
{L{\'o}pez-Valdivia} R.,  et~al., 2023, \mn@doi [\apj] {10.3847/1538-4357/acab04}, \href {https://ui.adsabs.harvard.edu/abs/2023ApJ...943...49L} {943, 49}

\bibitem[\protect\citeauthoryear{Luger \& Barnes}{Luger \& Barnes}{2015}]{luger_extreme_2015}
Luger R.,  Barnes R.,  2015, \mn@doi [Astrobiology] {10.1089/ast.2014.1231}, 15, 119

\bibitem[\protect\citeauthoryear{{Maehara}, {Shibayama}, {Notsu}, {Notsu}, {Honda}, {Nogami}  \& {Shibata}}{{Maehara} et~al.}{2015}]{maehara2015}
{Maehara} H.,  {Shibayama} T.,  {Notsu} Y.,  {Notsu} S.,  {Honda} S.,  {Nogami} D.,   {Shibata} K.,  2015, \mn@doi [Earth, Planets and Space] {10.1186/s40623-015-0217-z}, \href {https://ui.adsabs.harvard.edu/abs/2015EP&S...67...59M} {67, 59}

\bibitem[\protect\citeauthoryear{{Maschberger} \& {Kroupa}}{{Maschberger} \& {Kroupa}}{2009}]{maschberger}
{Maschberger} T.,  {Kroupa} P.,  2009, \mn@doi [\mnras] {10.1111/j.1365-2966.2009.14577.x}, \href {https://ui.adsabs.harvard.edu/abs/2009MNRAS.395..931M} {395, 931}

\bibitem[\protect\citeauthoryear{{McQuillan}, {Aigrain}  \& {Mazeh}}{{McQuillan} et~al.}{2013}]{mcquillan2013}
{McQuillan} A.,  {Aigrain} S.,   {Mazeh} T.,  2013, \mn@doi [\mnras] {10.1093/mnras/stt536}, \href {https://ui.adsabs.harvard.edu/abs/2013MNRAS.432.1203M} {432, 1203}

\bibitem[\protect\citeauthoryear{{Medina}, {Winters}, {Irwin}  \& {Charbonneau}}{{Medina} et~al.}{2020}]{medina2020}
{Medina} A.~A.,  {Winters} J.~G.,  {Irwin} J.~M.,   {Charbonneau} D.,  2020, \mn@doi [\apj] {10.3847/1538-4357/abc686}, \href {https://ui.adsabs.harvard.edu/abs/2020ApJ...905..107M} {905, 107}

\bibitem[\protect\citeauthoryear{{Medina}, {Winters}, {Irwin}  \& {Charbonneau}}{{Medina} et~al.}{2022}]{medina2022}
{Medina} A.~A.,  {Winters} J.~G.,  {Irwin} J.~M.,   {Charbonneau} D.,  2022, \mn@doi [\apj] {10.3847/1538-4357/ac77f9}, \href {https://ui.adsabs.harvard.edu/abs/2022ApJ...935..104M} {935, 104}

\bibitem[\protect\citeauthoryear{{Metcalfe} et~al.,}{{Metcalfe} et~al.}{2023}]{metcalfe2023}
{Metcalfe} T.~S.,  et~al., 2023, \mn@doi [\apjl] {10.3847/2041-8213/acce38}, 948, L6

\bibitem[\protect\citeauthoryear{{Miles-P{\'a}ez}, {Metchev}  \& {George}}{{Miles-P{\'a}ez} et~al.}{2023}]{milespaez2023}
{Miles-P{\'a}ez} P.~A.,  {Metchev} S.~A.,   {George} B.,  2023, \mn@doi [\mnras] {10.1093/mnras/stad273}, \href {https://ui.adsabs.harvard.edu/abs/2023MNRAS.521..952M} {521, 952}

\bibitem[\protect\citeauthoryear{{Mulders}, {Pascucci}  \& {Apai}}{{Mulders} et~al.}{2015}]{mulders2015}
{Mulders} G.~D.,  {Pascucci} I.,   {Apai} D.,  2015, \mn@doi [\apj] {10.1088/0004-637X/798/2/112}, \href {https://ui.adsabs.harvard.edu/abs/2015ApJ...798..112M} {798, 112}

\bibitem[\protect\citeauthoryear{{Murray} et~al.,}{{Murray} et~al.}{2022}]{murray}
{Murray} C.~A.,  et~al., 2022, \mn@doi [\mnras] {10.1093/mnras/stac1078}, \href {https://ui.adsabs.harvard.edu/abs/2022MNRAS.513.2615M} {513, 2615}

\bibitem[\protect\citeauthoryear{{Newton}, {Irwin}, {Charbonneau}, {Berta-Thompson}, {Dittmann}  \& {West}}{{Newton} et~al.}{2016}]{newton2016}
{Newton} E.~R.,  {Irwin} J.,  {Charbonneau} D.,  {Berta-Thompson} Z.~K.,  {Dittmann} J.~A.,   {West} A.~A.,  2016, \mn@doi [\apj] {10.3847/0004-637X/821/2/93}, \href {https://ui.adsabs.harvard.edu/abs/2016ApJ...821...93N} {821, 93}

\bibitem[\protect\citeauthoryear{{Newton}, {Irwin}, {Charbonneau}, {Berlind}, {Calkins}  \& {Mink}}{{Newton} et~al.}{2017}]{newton2017}
{Newton} E.~R.,  {Irwin} J.,  {Charbonneau} D.,  {Berlind} P.,  {Calkins} M.~L.,   {Mink} J.,  2017, \mn@doi [\apj] {10.3847/1538-4357/834/1/85}, \href {https://ui.adsabs.harvard.edu/abs/2017ApJ...834...85N} {834, 85}

\bibitem[\protect\citeauthoryear{{Nutzman} \& {Charbonneau}}{{Nutzman} \& {Charbonneau}}{2008}]{nutzman2008}
{Nutzman} P.,  {Charbonneau} D.,  2008, \mn@doi [\pasp] {10.1086/533420}, \href {https://ui.adsabs.harvard.edu/abs/2008PASP..120..317N} {120, 317}

\bibitem[\protect\citeauthoryear{{Paegert}, {Stassun}, {Collins}, {Pepper}, {Torres}, {Jenkins}, {Twicken}  \& {Latham}}{{Paegert} et~al.}{2021}]{paegert2021}
{Paegert} M.,  {Stassun} K.~G.,  {Collins} K.~A.,  {Pepper} J.,  {Torres} G.,  {Jenkins} J.,  {Twicken} J.~D.,   {Latham} D.~W.,  2021, \mn@doi [arXiv e-prints] {10.48550/arXiv.2108.04778}, \href {https://ui.adsabs.harvard.edu/abs/2021arXiv210804778P} {p. arXiv:2108.04778}

\bibitem[\protect\citeauthoryear{{Parker}}{{Parker}}{1955}]{parker1955}
{Parker} E.~N.,  1955, \mn@doi [\apj] {10.1086/146087}, \href {https://ui.adsabs.harvard.edu/abs/1955ApJ...122..293P} {122, 293}

\bibitem[\protect\citeauthoryear{Parsons et~al.,}{Parsons et~al.}{2018}]{parsons2018}
Parsons S.~G.,  et~al., 2018, \mn@doi [Monthly Notices of the Royal Astronomical Society] {10.1093/mnras/sty2345}, 481, 1083

\bibitem[\protect\citeauthoryear{{Pascucci} et~al.,}{{Pascucci} et~al.}{2016}]{pascucci2016}
{Pascucci} I.,  et~al., 2016, \mn@doi [\apj] {10.3847/0004-637X/831/2/125}, \href {https://ui.adsabs.harvard.edu/abs/2016ApJ...831..125P} {831, 125}

\bibitem[\protect\citeauthoryear{{Paudel}, {Gizis}, {Mullan}, {Schmidt}, {Burgasser}, {Williams}  \& {Berger}}{{Paudel} et~al.}{2018}]{paudel2018}
{Paudel} R.~R.,  {Gizis} J.~E.,  {Mullan} D.~J.,  {Schmidt} S.~J.,  {Burgasser} A.~J.,  {Williams} P. K.~G.,   {Berger} E.,  2018, \mn@doi [\apj] {10.3847/1538-4357/aab8fe}, \href {https://ui.adsabs.harvard.edu/abs/2018ApJ...858...55P} {858, 55}

\bibitem[\protect\citeauthoryear{{Pollack}, {Hubickyj}, {Bodenheimer}, {Lissauer}, {Podolak}  \& {Greenzweig}}{{Pollack} et~al.}{1996}]{pollack1996}
{Pollack} J.~B.,  {Hubickyj} O.,  {Bodenheimer} P.,  {Lissauer} J.~J.,  {Podolak} M.,   {Greenzweig} Y.,  1996, \mn@doi [\icarus] {10.1006/icar.1996.0190}, \href {https://ui.adsabs.harvard.edu/abs/1996Icar..124...62P} {124, 62}

\bibitem[\protect\citeauthoryear{{Quirrenbach} et~al.,}{{Quirrenbach} et~al.}{2018}]{quirrenbach2018}
{Quirrenbach} A.,  et~al., 2018, in {Evans} C.~J.,  {Simard} L.,   {Takami} H.,  eds,  Society of Photo-Optical Instrumentation Engineers (SPIE) Conference Series Vol. 10702, Ground-based and Airborne Instrumentation for Astronomy VII. p. 107020W, \mn@doi{10.1117/12.2313689}

\bibitem[\protect\citeauthoryear{{Raetz}, {Stelzer}  \& {Scholz}}{{Raetz} et~al.}{2020a}]{raetz2019}
{Raetz} S.,  {Stelzer} B.,   {Scholz} A.,  2020a, \mn@doi [Astronomische Nachrichten] {10.1002/asna.202013727}, \href {https://ui.adsabs.harvard.edu/abs/2020AN....341..519R} {341, 519}

\bibitem[\protect\citeauthoryear{{Raetz}, {Stelzer}, {Damasso}  \& {Scholz}}{{Raetz} et~al.}{2020b}]{raetz2020}
{Raetz} S.,  {Stelzer} B.,  {Damasso} M.,   {Scholz} A.,  2020b, \mn@doi [\aap] {10.1051/0004-6361/201937350}, \href {https://ui.adsabs.harvard.edu/abs/2020A&A...637A..22R} {637, A22}

\bibitem[\protect\citeauthoryear{{Raymond}, {Scalo}  \& {Meadows}}{{Raymond} et~al.}{2007}]{raymond2007}
{Raymond} S.~N.,  {Scalo} J.,   {Meadows} V.~S.,  2007, \mn@doi [\apj] {10.1086/521587}, \href {https://ui.adsabs.harvard.edu/abs/2007ApJ...669..606R} {669, 606}

\bibitem[\protect\citeauthoryear{{Rebull}, {Stauffer}, {Hillenbrand}, {Cody}, {Kruse}  \& {Powell}}{{Rebull} et~al.}{2022}]{rebull2022}
{Rebull} L.~M.,  {Stauffer} J.~R.,  {Hillenbrand} L.~A.,  {Cody} A.~M.,  {Kruse} E.,   {Powell} B.~P.,  2022, \mn@doi [\aj] {10.3847/1538-3881/ac75f1}, \href {https://ui.adsabs.harvard.edu/abs/2022AJ....164...80R} {164, 80}

\bibitem[\protect\citeauthoryear{{Reddy}, {Lambert}  \& {Allende Prieto}}{{Reddy} et~al.}{2006}]{reddy2006}
{Reddy} B.~E.,  {Lambert} D.~L.,   {Allende Prieto} C.,  2006, \mn@doi [\mnras] {10.1111/j.1365-2966.2006.10148.x}, \href {https://ui.adsabs.harvard.edu/abs/2006MNRAS.367.1329R} {367, 1329}

\bibitem[\protect\citeauthoryear{{Ricker} et~al.,}{{Ricker} et~al.}{2015}]{ricker2015}
{Ricker} G.~R.,  et~al., 2015, \mn@doi [Journal of Astronomical Telescopes, Instruments, and Systems] {10.1117/1.JATIS.1.1.014003}, \href {https://ui.adsabs.harvard.edu/abs/2015JATIS...1a4003R} {1, 014003}

\bibitem[\protect\citeauthoryear{{Rimmer}, {Xu}, {Thompson}, {Gillen}, {Sutherland}  \& {Queloz}}{{Rimmer} et~al.}{2018}]{rimmer2018}
{Rimmer} P.~B.,  {Xu} J.,  {Thompson} S.~J.,  {Gillen} E.,  {Sutherland} J.~D.,   {Queloz} D.,  2018, \mn@doi [Science Advances] {10.1126/sciadv.aar3302}, \href {https://ui.adsabs.harvard.edu/abs/2018SciA....4.3302R} {4, eaar3302}

\bibitem[\protect\citeauthoryear{Rodgers-Lee et~al.,}{Rodgers-Lee et~al.}{2023}]{rodgers-lee_energetic_2023}
Rodgers-Lee D.,  et~al., 2023, \mn@doi [Monthly Notices of the Royal Astronomical Society] {10.1093/mnras/stad900}, 521, 5880

\bibitem[\protect\citeauthoryear{{Rodr{\'\i}guez Mart{\'\i}nez}, {Lopez}, {Shappee}, {Schmidt}, {Jayasinghe}, {Kochanek}, {Auchettl}  \& {Holoien}}{{Rodr{\'\i}guez Mart{\'\i}nez} et~al.}{2020}]{rodriguez2020}
{Rodr{\'\i}guez Mart{\'\i}nez} R.,  {Lopez} L.~A.,  {Shappee} B.~J.,  {Schmidt} S.~J.,  {Jayasinghe} T.,  {Kochanek} C.~S.,  {Auchettl} K.,   {Holoien} T. W.~S.,  2020, \mn@doi [\apj] {10.3847/1538-4357/ab793a}, \href {https://ui.adsabs.harvard.edu/abs/2020ApJ...892..144R} {892, 144}

\bibitem[\protect\citeauthoryear{{Scalo} et~al.,}{{Scalo} et~al.}{2007}]{scalo2007}
{Scalo} J.,  et~al., 2007, \mn@doi [Astrobiology] {10.1089/ast.2006.0125}, \href {https://ui.adsabs.harvard.edu/abs/2007AsBio...7...85S} {7, 85}

\bibitem[\protect\citeauthoryear{{Scargle}}{{Scargle}}{1982}]{scargle1982}
{Scargle} J.~D.,  1982, \mn@doi [\apj] {10.1086/160554}, \href {https://ui.adsabs.harvard.edu/abs/1982ApJ...263..835S} {263, 835}

\bibitem[\protect\citeauthoryear{{Schaefer}, {King}  \& {Deliyannis}}{{Schaefer} et~al.}{2000}]{schaefer}
{Schaefer} B.~E.,  {King} J.~R.,   {Deliyannis} C.~P.,  2000, \mn@doi [\apj] {10.1086/308325}, \href {https://ui.adsabs.harvard.edu/abs/2000ApJ...529.1026S} {529, 1026}

\bibitem[\protect\citeauthoryear{{Schmidt}, {Hawley}, {West}, {Bochanski}, {Davenport}, {Ge}  \& {Schneider}}{{Schmidt} et~al.}{2015}]{schmidt2015}
{Schmidt} S.~J.,  {Hawley} S.~L.,  {West} A.~A.,  {Bochanski} J.~J.,  {Davenport} J. R.~A.,  {Ge} J.,   {Schneider} D.~P.,  2015, \mn@doi [\aj] {10.1088/0004-6256/149/5/158}, \href {https://ui.adsabs.harvard.edu/abs/2015AJ....149..158S} {149, 158}

\bibitem[\protect\citeauthoryear{{Sebastian} et~al.,}{{Sebastian} et~al.}{2021}]{sebastian2021}
{Sebastian} D.,  et~al., 2021, \mn@doi [\aap] {10.1051/0004-6361/202038827}, \href {https://ui.adsabs.harvard.edu/abs/2021A&A...645A.100S} {645, A100}

\bibitem[\protect\citeauthoryear{{Segura}}{{Segura}}{2018}]{segura2018}
{Segura} A.,  2018, in {Deeg} H.~J.,  {Belmonte} J.~A.,  eds, , Handbook of Exoplanets.
Springer Nature, p.~73, \mn@doi{10.1007/978-3-319-55333-7_73}

\bibitem[\protect\citeauthoryear{Segura, Walkowicz, Meadows, Kasting  \& Hawley}{Segura et~al.}{2010}]{segura_effect_2010}
Segura A.,  Walkowicz L.~M.,  Meadows V.,  Kasting J.,   Hawley S.,  2010, \mn@doi [Astrobiology] {10.1089/ast.2009.0376}, 10, 751

\bibitem[\protect\citeauthoryear{{Seli}, {Vida}, {Mo{\'o}r}, {P{\'a}l}  \& {Ol{\'a}h}}{{Seli} et~al.}{2021}]{seli2021}
{Seli} B.,  {Vida} K.,  {Mo{\'o}r} A.,  {P{\'a}l} A.,   {Ol{\'a}h} K.,  2021, \mn@doi [\aap] {10.1051/0004-6361/202040098}, \href {https://ui.adsabs.harvard.edu/abs/2021A&A...650A.138S} {650, A138}

\bibitem[\protect\citeauthoryear{{Serna} et~al.,}{{Serna} et~al.}{2021}]{serna2021}
{Serna} J.,  et~al., 2021, \mn@doi [\apj] {10.3847/1538-4357/ac300a}, \href {https://ui.adsabs.harvard.edu/abs/2021ApJ...923..177S} {923, 177}

\bibitem[\protect\citeauthoryear{{Shields}, {Ballard}  \& {Johnson}}{{Shields} et~al.}{2016}]{shieldsetal2016}
{Shields} A.~L.,  {Ballard} S.,   {Johnson} J.~A.,  2016, \mn@doi [\physrep] {10.1016/j.physrep.2016.10.003}, \href {https://ui.adsabs.harvard.edu/abs/2016PhR...663....1S} {663, 1}

\bibitem[\protect\citeauthoryear{{Silverberg}, {Kowalski}, {Davenport}, {Wisniewski}, {Hawley}  \& {Hilton}}{{Silverberg} et~al.}{2016}]{silverberg2016}
{Silverberg} S.~M.,  {Kowalski} A.~F.,  {Davenport} J. R.~A.,  {Wisniewski} J.~P.,  {Hawley} S.~L.,   {Hilton} E.~J.,  2016, \mn@doi [\apj] {10.3847/0004-637X/829/2/129}, \href {https://ui.adsabs.harvard.edu/abs/2016ApJ...829..129S} {829, 129}

\bibitem[\protect\citeauthoryear{{Skumanich}}{{Skumanich}}{1972}]{skumanich1972}
{Skumanich} A.,  1972, \mn@doi [\apj] {10.1086/151310}, \href {https://ui.adsabs.harvard.edu/abs/1972ApJ...171..565S} {171, 565}

\bibitem[\protect\citeauthoryear{{Stassun} et~al.,}{{Stassun} et~al.}{2019}]{stassun2019}
{Stassun} K.~G.,  et~al., 2019, \mn@doi [\aj] {10.3847/1538-3881/ab3467}, \href {https://ui.adsabs.harvard.edu/abs/2019AJ....158..138S} {158, 138}

\bibitem[\protect\citeauthoryear{{Sullivan} et~al.,}{{Sullivan} et~al.}{2015}]{sullivan2015}
{Sullivan} P.~W.,  et~al., 2015, \mn@doi [\apj] {10.1088/0004-637X/809/1/77}, \href {https://ui.adsabs.harvard.edu/abs/2015ApJ...809...77S} {809, 77}

\bibitem[\protect\citeauthoryear{{Tamburo} et~al.,}{{Tamburo} et~al.}{2022}]{tamburo2022}
{Tamburo} P.,  et~al., 2022, \mn@doi [\aj] {10.3847/1538-3881/ac64aa}, \href {https://ui.adsabs.harvard.edu/abs/2022AJ....163..253T} {163, 253}

\bibitem[\protect\citeauthoryear{{Tannock} et~al.,}{{Tannock} et~al.}{2021}]{tannock2021}
{Tannock} M.~E.,  et~al., 2021, \mn@doi [\aj] {10.3847/1538-3881/abeb67}, \href {https://ui.adsabs.harvard.edu/abs/2021AJ....161..224T} {161, 224}

\bibitem[\protect\citeauthoryear{{Tian} et~al.,}{{Tian} et~al.}{2015}]{tian2015}
{Tian} H.-J.,  et~al., 2015, \mn@doi [\apj] {10.1088/0004-637X/809/2/145}, \href {https://ui.adsabs.harvard.edu/abs/2015ApJ...809..145T} {809, 145}

\bibitem[\protect\citeauthoryear{Tilley, Segura, Meadows, Hawley  \& Davenport}{Tilley et~al.}{2019}]{tilley_modeling_2019}
Tilley M.~A.,  Segura A.,  Meadows V.,  Hawley S.,   Davenport J.,  2019, \mn@doi [Astrobiology] {10.1089/ast.2017.1794}, 19, 64

\bibitem[\protect\citeauthoryear{{Tu}, {Yang}, {Zhang}  \& {Wang}}{{Tu} et~al.}{2020}]{tu2020}
{Tu} Z.-L.,  {Yang} M.,  {Zhang} Z.~J.,   {Wang} F.~Y.,  2020, \mn@doi [\apj] {10.3847/1538-4357/ab6606}, \href {https://ui.adsabs.harvard.edu/abs/2020ApJ...890...46T} {890, 46}

\bibitem[\protect\citeauthoryear{{Wheatland}}{{Wheatland}}{2004}]{wheatland2004}
{Wheatland} M.~S.,  2004, \mn@doi [\apj] {10.1086/421261}, \href {https://ui.adsabs.harvard.edu/abs/2004ApJ...609.1134W} {609, 1134}

\bibitem[\protect\citeauthoryear{{Wright}, {Newton}, {Williams}, {Drake}  \& {Yadav}}{{Wright} et~al.}{2018}]{wright2018}
{Wright} N.~J.,  {Newton} E.~R.,  {Williams} P. K.~G.,  {Drake} J.~J.,   {Yadav} R.~K.,  2018, \mn@doi [\mnras] {10.1093/mnras/sty1670}, \href {https://ui.adsabs.harvard.edu/abs/2018MNRAS.479.2351W} {479, 2351}

\bibitem[\protect\citeauthoryear{{Yang} et~al.,}{{Yang} et~al.}{2017}]{yang2017}
{Yang} H.,  et~al., 2017, \mn@doi [\apj] {10.3847/1538-4357/aa8ea2}, \href {https://ui.adsabs.harvard.edu/abs/2017ApJ...849...36Y} {849, 36}

\bibitem[\protect\citeauthoryear{{Yang}, {Zhang}, {Meng}, {Han}, {Misra}, {Yang}  \& {Pi}}{{Yang} et~al.}{2023}]{yang2023}
{Yang} Z.,  {Zhang} L.,  {Meng} G.,  {Han} X.~L.,  {Misra} P.,  {Yang} J.,   {Pi} Q.,  2023, \mn@doi [\aap] {10.1051/0004-6361/202142710}, \href {https://ui.adsabs.harvard.edu/abs/2023A&A...669A..15Y} {669, A15}

\bibitem[\protect\citeauthoryear{{Zechmeister} et~al.,}{{Zechmeister} et~al.}{2019}]{zechmeister2019}
{Zechmeister} M.,  et~al., 2019, \mn@doi [\aap] {10.1051/0004-6361/201935460}, \href {https://ui.adsabs.harvard.edu/abs/2019A&A...627A..49Z} {627, A49}

\bibitem[\protect\citeauthoryear{{do Amaral}, {Barnes}, {Segura}  \& {Luger}}{{do Amaral} et~al.}{2022}]{doamaral2022}
{do Amaral} L. N.~R.,  {Barnes} R.,  {Segura} A.,   {Luger} R.,  2022, \mn@doi [\apj] {10.3847/1538-4357/ac53af}, \href {https://ui.adsabs.harvard.edu/abs/2022ApJ...928...12D} {928, 12}

\makeatother
\end{thebibliography}









\twocolumn


\bsp	
\label{lastpage}
\end{document}